\documentclass[acmsmall,screen,nonacm]{acmart}

\usepackage{graphicx} 
\usepackage{amsmath}
\usepackage{amsthm}
\usepackage{thmtools}

\usepackage{amssymb}
\usepackage{tikz}
\usepackage{booktabs}
\usepackage{multirow}
\usepackage{colortbl}
\usepackage{array}
\usepackage{xspace}
\usepackage{soul}
\usepackage{stmaryrd}
\usepackage{subcaption}
\usepackage{enumitem}
\usepackage{hyperref}
\usepackage{nicematrix}
\usepackage{tikz}
\usepackage{listings}
\usepackage{wrapfig}
\usepackage{microtype}
\usetikzlibrary{matrix, arrows.meta, positioning}
\usetikzlibrary{shapes.geometric, arrows.meta, calc, fit, backgrounds}
\definecolor{Burgundy}{RGB}{158,5,8}
\definecolor{Cerulean}{rgb}{0.0, 0.48, 0.65}

\newcommand{\todo}[1]{{\color{Cerulean} #1}
}

\newcommand{\rev}[1]{#1}
\theoremstyle{plain}
\newtheorem{theorem}{Theorem}[section]
\newtheorem{lemma}[theorem]{Lemma}
\newtheorem{proposition}[theorem]{Proposition}

\theoremstyle{definition}

\newtheorem{innerexample}{Example}

\newenvironment{example}
  {\pushQED{\qed}%
   \innerexample}
  {\popQED\endinnerexample}

\theoremstyle{remark}

\allowdisplaybreaks
\newcommand{\rec}[0]{\textit{Rec}}

\newcommand{\zk}[1]{\ensuremath{z^{#1}}}

\sethlcolor{yellow}

\newcommand{\vecrec}[0]{\textsf{VecRec}\xspace}

\newcommand{\CC}{C\nolinebreak\hspace{-.05em}\raisebox{.4ex}{\tiny\bf +}\nolinebreak\hspace{-.10em}\raisebox{.4ex}{\tiny\bf +}}
\def\CC{{C\nolinebreak[4]\hspace{-.05em}\raisebox{.4ex}{\tiny\bf ++}}}
\def\compactsub{\!-\!}

\newcommand{\order}[0]{\textit{ord}}

\newcommand{\semof}[1]{\left\llbracket #1\right\rrbracket^\sigma}

\def\algone{\ensuremath{\boldsymbol{\mathsf{1}}_{\tau}}}
\def\algzero{\ensuremath{\boldsymbol{\mathsf{0}}_{\tau}}}
\def\algplus{\ensuremath{\oplus_\tau}}
\def\algtimes{\ensuremath{\otimes_\tau}}
\def\algminus{\ensuremath{\ominus_\tau}}
\def\vecn{\ensuremath{\mathbf{n}}}
\def\veczero{\ensuremath{\mathbf{0}}}
\def\vecone{\ensuremath{\mathbf{1}}}
\def\veci{\ensuremath{\mathbf{i}}}
\def\veck{\ensuremath{\mathbf{k}}}

\def\zerokernel{\ensuremath{\mathsf{0}}}

\begin{abstract}
Recurrences are ubiquitous in signal and image processing, bioinformatics, 
numerical computing, combinatorial optimization, and many other domains. 
Because recurrences seem inherently sequential,
  they are often implemented serially
  and thus do not leverage vector instructions in modern hardware.
This leaves much of the available,
  often highly parallel, hardware underutilized.

In this paper,
  we present an algebra of recurrences over semirings and
  study algebraic identities that enable vectorization.
Using our algebraic framework, 
 we present a number of vectorizing transformations that yield efficient, vectorized implementations across various settings.
These transformations naturally extend to 2D and mutually recursive recurrences.

We implement the recurrence algebra and the vectorizing transformations into a compiler called \vecrec.
\vecrec compiles recurrence programs to efficient vectorized \CC{} implementations.
The generated kernels achieve up to an order of magnitude speedup over serial implementations across a wide variety of recurrences,
 and are faster than existing libraries with manually-crafted vectorization in signal processing and bioinformatics.
\end{abstract}

\setcopyright{cc}
\setcctype{by}
\acmDOI{10.1145/3871805}
\acmYear{2027}
\acmJournal{PACMPL}
\acmVolume{11}
\acmNumber{POPL}
\acmArticle{103}
\acmMonth{1}
\acmSubmissionID{popl27main-p1234-p}
\received{2026-07-09}
\received[accepted]{2026-10-05}



\begin{document}
\title{Vectorizing Recurrences over Semirings}

\author{Yihong Zhang}
\correspondingauthor
\orcid{0009-0006-5928-4396}
\affiliation{%
  \institution{University of Washington}
  \city{Seattle}
  \country{USA}
}
\email{yz489@cs.washington.edu}

\author{Bala Vinaithirthan}
\orcid{0000-0002-0829-4360}
\affiliation{%
  \institution{Stanford University}
  \city{Stanford}
  \country{USA}
}
\email{bala3@stanford.edu}

\author{Karima Ma}
\orcid{0000-0003-4180-6433}
\affiliation{%
  \institution{Adobe}
  \city{New York}
  \country{USA}
}
\email{kma@adobe.com}

\author{Zachary Tatlock}
\orcid{0000-0002-4731-0124}
\affiliation{%
  \institution{University of Washington}
  \city{Seattle}
  \country{USA}
}
\email{ztatlock@cs.washington.edu}

\author{Maaz Bin Safeer Ahmad}
\orcid{0000-0002-8113-7580}
\affiliation{%
  \institution{Adobe}
  \city{Seattle}
  \country{USA}
}
\email{mahmad@adobe.com}

\author{Andrew Adams}
\orcid{0000-0002-6766-670X}
\affiliation{%
  \institution{Adobe}
  \city{San Francisco}
  \country{USA}
}
\email{andrew.b.adams@gmail.com}

\renewcommand{\shortauthors}{Y.~Zhang, B.~Vinaithirthan, K.~Ma, Z.~Tatlock, M.~B.~S.~Ahmad, and A.~Adams}

\maketitle

\section{Introduction}
\label{sec:intro}

A recurrence defines a sequence of values, where each value $f(x)$ depends on previous values (e.g., $f(x\compactsub 1)$).
Recurrences are ubiquitous in many computational domains. 
The simplest form of a recurrence is a prefix sum of an input sequence $g$:
 {\small \[ \tag*{(Prefix Sum)} f(x) = f(x\compactsub 1) + g(x).
 \quad\quad\quad\quad\quad\quad\quad\quad\quad\quad\quad\quad
 \quad\quad\quad\quad\quad\quad\quad\quad\quad\quad \]}

In signal processing, many classical filters can be expressed as recurrences. They are known as infinite-impulse response (IIR) filters. For example,
{\small \[ \tag*{(A Band-Pass Filter)}f(x) = 1.8 \cdot f(x\compactsub1) - 0.9 \cdot f(x\compactsub2) + g(x).
\quad\quad\quad\quad\quad\quad\quad\quad\quad\quad\quad\quad\]}

In computer graphics, 
 an approximate distance transform is a recurrence with $\min$ and $+$:
{\small \[
\tag*{(Distance Transform)}
f(x, y) = \min(f(x\compactsub1, y) + 1, f(x, y\compactsub1) + 1, f(x\compactsub1, y\compactsub1) + \sqrt{2}, g(x, y)).
\quad\ 
\]}

Sequence alignment algorithms in bioinformatics have a similar pattern,
 where $s$ is a score matrix and $c_\text{gap}$ is a constant gap penality:
{\small\[
\tag*{(Needleman--Wunsch)}
f(x,y) = \max( f(x\compactsub1,y\compactsub1) + s(x,y), f(x\compactsub1,y) + c_\text{gap}, f(x,y\compactsub1) + c_\text{gap} ).
\]}

Even dynamic programming problems on LeetCode often reduce to computing recurrences:
{\small
\[
\tag*{(LeetCode 188.~Buy/Sell Stock)}
\begin{aligned}
f_1(x, y) = \max(f_1(x, y\compactsub1), f_2(x\compactsub1, y\compactsub1) - g(y)),\\
f_2(x, y) = \max(f_1(x, y\compactsub1) + g(y),\;\! \ \ \ \ f_2(x, y\compactsub1)).
\end{aligned}
\quad\quad\quad\quad \quad\quad\ \ 
\]
}


These tasks from signal processing, graphics, bioinformatics, and LeetCode
are all instances of the same problem: computing an $F$ that satisfies
\[F = A\otimes F \oplus G\] over a semiring $\tau$. In this definition,
  $A$ is a convolution \textit{kernel}
  (a finite array of coefficients, or \textit{taps}),
  $G$ is the input \textit{signal}
  (a stream of values), 
  $F$ is the output signal,
 and convolution $\otimes$ computes
  a sliding weighted sum of the signal by the kernel\footnote{The convolution $\otimes$ in this paper should not be confused with matrix multiplication.
  \rev{Intuitively, $(A\otimes F)(\vecn)={\bigoplus_\tau}_{\veci} A(\veci)\otimes_\tau F(\vecn\compactsub\veci)$,
  where $\oplus_\tau$ and $\otimes_\tau$ are semiring operations.}}.
For example, by instantiating the recurrence over the arithmetic (semi)ring and letting $A$ be the kernel $[0, 1.8, -0.9]$, we obtain the band-pass filter above. Similarly, if we use the $(\min,+)$-semiring and set the (2D) kernel to $\left[\begin{smallmatrix}
    \infty, & 1\\
    1, & \sqrt{2}
\end{smallmatrix}\right]$, we obtain the distance transform.

These recurrence relations are fundamental building blocks in their respective domains. Computing them efficiently is important, yet difficult.
While modern vector instruction sets like AVX-512 allow computing 16 values at a time,
 these recurrences enforce dependencies on previous output values (e.g., $f(x)$ depends on $f(x\compactsub 1)$),
 forcing a serial implementation.

Previous work has studied different aspects of speeding up these recurrences,
 such as multithreading~\citep{recfilter,jump-flooding}, tiling~\citep{reptile}, and data layout transformations~\citep{striped}.
This work takes a different approach:
 we vectorize recurrences through algebraic transformations.
Unlike prior approaches that are tailored to specific application domains,
 our approach is general:
 we can vectorize a wide class of recurrences, including all those in our examples,
 with speedups of up to an order of magnitude.
Core to our approach are the algebraic structures of the recurrences $F=A\otimes F\oplus G$.
We define an algebra for recurrences parametrized by a semiring $\tau$.
This algebra abstracts over the orders of recurrences\footnote{\rev{The order of a recurrence is how far back it looks (\autoref{sec:semantics}). For example, prefix sum is a first-order recurrence.}},
 the numerical values of coefficients,
 and the concrete operators, retaining only their algebraic properties.
We then present algebraic identities for recurrences over different semirings
 that enable vectorization for a wide range of real-world recurrences.

We instantiate the recurrence algebra and vectorizing transformations in a compiler dubbed \vecrec.
Given an input expression,
 \vecrec exhaustively applies the transformation schemes to derive a set of vectorization-friendly alternative forms.
It generates an efficient vectorized \CC{} kernel for each form and picks the best-performing kernel based on runtime.
We evaluate \vecrec{} on a variety of recurrences from different domains, many of which have not been vectorized before.
The results show \vecrec{}-generated kernels are significantly faster than unvectorized implementations and, when vectorized baselines are available, outperform the fastest implementations with handcrafted vectorization.

\section{Overview}

To convey the vectorizing transformations this paper considers,
we first (painfully) vectorize a recurrence by manual term manipulation.
We then sketch our recurrence algebra and show how a handful of
simple identities can transform the same example into the same vectorized form.

\begin{figure}[t]
\centering
\resizebox{\textwidth}{!}{%
\begin{tikzpicture}[
    font=\small,
    >=Stealth,
    region/.style={draw, rounded corners=3pt, inner sep=5pt},
    regtitle/.style={font=\small, inner sep=0pt},
    tech/.style={font=\small, inner sep=0pt},
    leaf/.style={draw, rounded corners=3pt, fill=gray!10, align=center, inner xsep=5pt, inner ysep=4pt},
    src/.style={draw, rounded corners=2pt, align=center, inner sep=3pt, minimum height=6mm, minimum width=14mm},
    arr/.style={->, thick},
    arrlbl/.style={font=\footnotesize, align=center, inner sep=1pt}
]
\node[leaf] (ring) {%
  \textbf{Ring} (e.g., arithmetic)\\[1pt]
  Dilation \& delay};
\node[leaf, right=5mm of ring] (idem) {%
  \textbf{Idempotent semiring} (e.g., tropical)\\[1pt]
  Dilation via idempotence};

\node[regtitle, anchor=south west] (midtitle)
  at ([xshift=-5pt, yshift=2mm]ring.north west) {\textbf{Time-invariant kernel} (\autoref{sec:dilation})};
\node[region, draw=none, fit=(ring)(idem)(midtitle)] (mid) {};

\node[tech, anchor=south west] (gen)
  at ([yshift=3mm]mid.north west)
  {Fast associative scan \;$\cdot$\; Path-based decomposition};
\node[regtitle, anchor=south west] (outtitle)
  at ([yshift=2mm]gen.north west) {\textbf{Any kernel} (\autoref{sec:general-recurrences})};
\begin{scope}[on background layer]
  \node[region, fill=white, fit=(mid)(gen)(outtitle)] (outer) {};
  \node[region, fill=gray!4, fit=(ring)(idem)(midtitle)] {};
\end{scope}

\coordinate (inTop) at ([yshift=7mm]outer.west);
\coordinate (inBot) at ([yshift=-7mm]outer.west);
\node[src, left=24mm of inTop] (twod) {2D\\recurrence};
\node[src, left=24mm of inBot] (mut)  {Mutual\\recurrence};
\draw[arr] (twod.east) -- node[arrlbl, above=1pt] {marginalization} node[arrlbl, below=1pt] {(\autoref{sec:extension})} (inTop);
\draw[arr] (mut.east)  -- node[arrlbl, above=1pt] {lifting to\\kernel matrices} node[arrlbl, below=1pt] {(\autoref{sec:extension})} (inBot);
\end{tikzpicture}
}
\caption{Vectorization techniques proposed in this paper.
Fast associative scan and path-based decomposition apply to any recurrence,
 while dilation requires a time-invariant kernel over a ring or an idempotent semiring.
\vecrec{} use all applicable transformations and use profiling to find the fastest kernel.}
\label{fig:decision-diagram}
\end{figure}

\subsection{Vectorizing a Second-Order Digital Filter}
\label{sec:overview-example}
Consider the following band-pass filter again:
{\small
\begin{align}
f(x) = a f(x\compactsub 1) + b f(x\compactsub 2) + g(x)\quad \quad a=1.8, b=-0.9.
\label{so-iir}
\end{align}
}

As written, this recurrence is not directly vectorizable: the data dependence of $f(x)$ on $f(x-1)$ prevents computing consecutive outputs in a single SIMD vector.

To break this dependence, we expand $f(x\compactsub 1)$ and simplify:
{\small
\begin{align}
f(x) = (a^2+b)f(x\compactsub 2) + ab f(x\compactsub 3) + g(x) + ag(x\compactsub 1).
\label{so-iir-delay}
\end{align}
}
After this substitution,
  $f(x)$ no longer depends on $f(x\compactsub 1)$,
  so we can compute two consecutive outputs in a single vector,
  as the following ``vectorized'' pseudocode\footnote{
    Here \texttt{f[x:x+2]} is the length-two vector of 
    consecutive outputs (half-open, Python-style slice notation), \texttt{*} is elementwise multiplication with
    scalars broadcast over vectors, and
    \texttt{+} is elementwise addition.
  } shows:
{\small
\begin{verbatim}
f[x:x+2] = (a*a+b) * f[x-2:x] + a*b * f[x-3:x-1] + g[x:x+2] + a * g[x-1:x+1]
\end{verbatim}
}
Repeating this process for $f(x\compactsub 2)$, $f(x\compactsub 3)$, \ldots, $f(x\compactsub n)$
 lets us compute $n$ elements in parallel.

This approach, however, creates too much extra work.
Each substitution lets us compute one more element in parallel, but it also incurs one fused multiply-add (fma) with a non-recursive $g$ term (e.g., $g(x\compactsub 1)$).
The original recurrence takes $2$ fmas to compute one element,
while the $n$-times-substituted version takes $n+2$ vectorized fmas to compute $n+1$ elements in parallel, so we reduce the number of fmas used by 
{\small $(2/1)\big/\bigl((n+2)/(n+1)\bigr)\leq 2$ times}.

To do better, observe that by rearranging \autoref{so-iir},
$f(x\compactsub 1)$ can be defined using $f(x)$ and $f(x\compactsub2)$:
{\small
\begin{align}
    f(x\compactsub 1)= \dfrac{f(x)-bf(x\compactsub 2)-g(x)}{a}
    \label{so-iir-x-1}
\end{align}
}
Using this to eliminate $f(x\compactsub 3)$ in \autoref{so-iir-delay}, we obtain a definition of $f(x)$ in terms of $f(x\compactsub 2)$ and $f(x\compactsub 4)$,
{\small
\begin{align}
    f(x) &= (a^2+b)f(x\compactsub 2) + ab \left(\dfrac{f(x\compactsub 2)-bf(x\compactsub 4)-g(x\compactsub 2)}{a}\right) + g(x) + ag(x\compactsub 1)\\
    &= a' f(x\compactsub 2)+b'f(x\compactsub 4)+g'(x)\label{so-iir-dilated}\\
    &\quad\quad\quad\text{where }\quad a'=a^2+2b,\quad b'=-b^2,\quad g'(x)=g(x) + ag(x\compactsub 1) - bg(x\compactsub 2).
\end{align}
}
%
Computing the auxiliary $g'$ costs 2 additional fmas.
We can apply the trick again to $f(x)=a' f(x\compactsub 2)+b'f(x\compactsub 4)+g'(x)$: expand $f(x\compactsub 2)$, define $f(x\compactsub 6)$ in terms of $f(x\compactsub 4)$ and $f(x\compactsub 8)$, and substitute.
We now define $f(x)$ in terms of $f(x\compactsub 4)$ and $f(x\compactsub 8)$,
 moving the self-dependency from 2 to 4 elements away.
In effect, this procedure doubles the distances between a recurrence's coefficients. We call it \textit{dilation}.

By repeatedly applying dilation, we obtain a tower of definitions:
{\small
\begin{align*}
    g_1(x) &= a_0\quad g(x\compactsub 1)\quad \quad\! +b_0g(x\compactsub2) \quad\quad\ + g(x),\\
    &\ldots\\
    g_n(x) &= a_{n\compactsub1}g_{n\compactsub1}(x\compactsub2^{n\compactsub1}) + b_{n\compactsub1}g_{n\compactsub 1}(x\compactsub2^n)+ g_{n\compactsub1}(x),\\
    f(x) &= a_n \ \ \  f(x\compactsub2^n)\quad\ \ + b_n f(x\compactsub2^{n+1}) \ \ \ + g_n(x).
\end{align*}
}
Each layer of the tower doubles the distance between coefficients. \rev{Moreover, since the definition of each $g_i$ does not depend on $g_i$ itself, we can compute the elements of $g_i$ in parallel, for any $i=1,\ldots, n$.}
In total, this tower of definitions requires $2\cdot (n+1)$ vectorized fmas, but it can compute $2^n$ elements at once.
In other words, $n$ steps of dilation reduce the number of fmas by $\dfrac{2/1}{2\cdot(n+1)/2^n}\times = \dfrac{2^n}{n+1}\times$.

This trick is known as scattered-lookahead \cite{scattered-lookahead} 
in digital circuit design but is largely overlooked in software,
 where it can yield significant speedups, as we will show:
 on an AMD Zen 5 CPU,
  dilating the above filter four times achieves a $4.86\times$ speedup.

In this work,
 we systematically study such vectorization techniques.
Dilation for arithmetic recurrences is just one of many transformations that enable vectorization;
 \autoref{fig:decision-diagram} organizes the full set used in this paper
 by the algebraic property each exploits.

\subsection{The Recurrence Algebra}

We propose an algebra of linear, fixed-order recurrences, inspired by the Z-transform~\citep{oppenheim1999discrete}.
Central to our recurrence algebra are two types of values, \textit{kernels} and \textit{signals}, and two types of operations, \textit{convolution} ($\otimes$) and (\textit{pointwise}) \textit{addition} ($\oplus$).

Under this algebra, a recurrence $F$ is defined as a signal that satisfies
\(
F=A\otimes F\oplus G
\).
Here,
$G$ and $F$ are the input and output signals, and $A$ is a convolution kernel.
That is, $F$ is defined as the convolution of itself with a kernel $A$, plus an external input $G$.
To avoid self-dependency (i.e., $f(x)$ depending on $f(x)$), we require the kernel $A$ to be \emph{causal}: it is zero at the origin, i.e.,
$A[\veczero]=\algzero$.

The recurrence algebra is parametrized by the semiring it operates over. 
For instance,
 we can also interpret $F=A\otimes F\oplus G$ over the tropical semiring $(\mathbb{R}\cup\{\infty\}, \min, +, \infty, 0)$,
 where $\oplus$ is minimum and $\otimes$ is addition.
Tropical recurrences capture DP-style optimization problems
 such as the distance transform in image processing and sequence alignment in bioinformatics.

\subsection{Identities of the Recurrence Algebra}

The identities of the recurrence algebra express the vectorizing transformations in \autoref{sec:overview-example} cleanly.
For instance, arithmetic is not just a semiring but a ring, so it has an additive inverse (subtraction), denoted as $\ominus$. This allows us to perform the following derivation.

Given a time-invariant arithmetic kernel $A=[0, a_1, a_2, \ldots]$, let us define
$E$ and $O$ as the even and odd terms of $A$, i.e.,
 $E = [0, 0, a_2, 0, a_4, \ldots]$ and $O=[0, a_1, 0, a_3,\ldots]$,
 such that $A=E\oplus O$. We have
{\small
\begin{align*}
    F&=A\otimes F\oplus G\\
    (I\ominus A)\otimes F&=G\\
    (I\ominus E\ominus O)\otimes F&=G\\
    (I\ominus E\oplus O)\otimes (I\ominus E\ominus O)\otimes F&= (I\ominus E\oplus O)\otimes G\\
    ((I\ominus E)^2\ominus O^2)\otimes F&=(I\ominus E\oplus O)\otimes G\\
    F&=(\ominus E^2\oplus E\oplus E\oplus O^2)\otimes F \oplus  (I\ominus E\oplus O)\otimes G
\end{align*}
}
where $I$ is the identity kernel $I=[1]$.
This is exactly the dilation of \autoref{sec:overview-example}: for $A=[0, a, b], E=[0, 0, b], O=[0, a]$, the new kernel $\ominus E^2\oplus E\oplus E\oplus O^2=[0, 0, 0, 0, -b^2] + [0, 0, 2b] + [0, 0, a^2]=[0, 0, a^2+2b, 0, -b^2]$ matches the coefficients in \autoref{so-iir-dilated},
but the algebraic form generalizes to kernels with any number of taps.

Similarly, we derive vectorizing transformations for idempotent, time-varying,
 and mutual recurrences from identities in our algebraic framework.

\subsection{A Vectorizing Compiler for Recurrences}

We instantiate the recurrence algebra and 
 the vectorizing transformations in a compiler, dubbed \vecrec.
Given an input recurrence, \vecrec automatically
 synthesizes several equivalent, vectorization-friendly forms,
 emits efficient vectorized \CC{} code for each form,
 and selects the best-performing form by throughput.
The generated kernel is efficient and pluggable into existing systems.

\subsection{Paper Organization}

The rest of the paper is organized as follows.
\autoref{sec:syntax-semantics} introduces the syntax and semantics of the recurrence algebra
 and discusses the properties of $\oplus$ and $\otimes$.
\autoref{sec:algebra} discusses the algebraic identities of the recurrence operator.
\autoref{sec:vectorization-tricks} presents the vectorizing transformations
 of recurrences and extensions to two-dimensional recurrences and mutual recurrences.
\autoref{sec:compiler} describes our compiler implementation
 that generates vectorized kernels for recurrences.
\autoref{sec:evaluation} evaluates our compiler on a diverse range of recurrences and
 presents a case study on bioinformatic recurrences.

\definecolor{headerblue}{RGB}{220,230,242}
\definecolor{rowlightA}{RGB}{237,245,253}
\definecolor{rowlightB}{RGB}{245,250,245}

\section{Syntax and Semantics}
\label{sec:syntax-semantics}

This section formally defines the recurrence algebra,
 which captures convolutions and (linear) recurrences over different semirings
 and lays the foundation for our transformations.
After defining its syntax and semantics, we present the properties of its non-recursive fragment.

\subsection{Semirings and Rings}
The recurrence algebra is parametrized by a commutative semiring $\tau$.
A \emph{semiring} is a tuple
$(\tau, \algplus, \algtimes, \algzero, \algone)$
such that $(\tau, \algplus, \algzero)$ is a commutative monoid,
$(\tau, \algtimes, \algone)$ is a monoid,
$\algtimes$ distributes over $\algplus$ on both sides,
and $\algzero$ is absorbing for $\algtimes$; that is:
\[
  \small
  \renewcommand{\arraystretch}{1.2}
  \begin{array}{r@{\quad}l@{\qquad}l}
    \textit{identity:}
      & \algzero\algplus a=a\algplus \algzero=a
      & \algone\algtimes a=a\algtimes \algone=a\\
    \textit{associativity:}
      & (a\algplus b)\algplus c=a\algplus(b\algplus c)
      & (a\algtimes b)\algtimes c=a\algtimes(b\algtimes c)\\
    \textit{commutativity:}
      & a\algplus b=b\algplus a &\\
    \textit{distributivity:}
      & (a\algplus b)\algtimes c = a\algtimes c\algplus b\algtimes c
      & a\algtimes (b\algplus c) = a\algtimes b\algplus a\algtimes c\\
    \textit{annihilation:}
      & \algzero \algtimes a = a \algtimes \algzero = \algzero &
  \end{array}
\]

An \emph{idempotent semiring}, also known as a \emph{dioid}, is a semiring that satisfies $a\algplus a=a$.
A \emph{commutative semiring} is a semiring that satisfies $a\algtimes b=b\algtimes a$.
A \emph{ring} is a semiring in which every element $a$ has an additive inverse $\algminus a$
 with $a\algplus (\algminus a)=\algzero$.

\begin{example}[Arithmetic ring]
The real numbers $(\mathbb{R}, +, \times, 0, 1)$ form a commutative ring,
with $\algplus = +$, $\algtimes = \times$, $\algzero = 0$, $\algone = 1$,
and additive inverse $\algminus a = -a$.
Recurrences over the arithmetic ring capture many recursive filters in signal processing.
\end{example}

\begin{example}[Tropical semiring]
The min-plus tropical semiring is
$(\mathbb{R}\cup\{+\infty\}, \min, +, +\infty, 0)$,
with $\algplus = \min$, $\algtimes = +$, $\algzero = +\infty$,
and $\algone = 0$.
Addition is idempotent ($\min(a, a) = a$), so this is an idempotent
commutative semiring.
It is not a ring because $\min$ has no inverse.
The tropical semiring appears in many combinatorial optimization problems, e.g., in bioinformatics.
Similarly, the max-plus tropical semiring is defined as $(\mathbb{R}\cup\{-\infty\}, \max, +, -\infty, 0)$.
\end{example}

A \emph{(left) semimodule} over a semiring
 $(\tau, \algplus, \algtimes, \algzero, \algone)$
 consists of a set $M$ with an addition $\oplus_M$, a zero $\boldsymbol{0}_M$,
 and a \emph{scalar multiplication} $\otimes_M:\tau\times M\to M$
 such that $(M, \oplus_M, \boldsymbol{0}_M)$ is a commutative monoid
 and the scalar multiplication is compatible with the semiring structure;
 that is, for all $r, s\in\tau$ and $m, n, p\in M$:
\[
  \small
  \renewcommand{\arraystretch}{1.2}
  \begin{array}{r@{\quad}l@{\qquad}l}
    \textit{identity:}
      & \boldsymbol{0}_M\oplus_M m=m\oplus_M\boldsymbol{0}_M=m
      & \algone\otimes_M m=m\\
    \textit{associativity:}
      & (m\oplus_M n)\oplus_M p=m\oplus_M(n\oplus_M p)
      & (r\algtimes s)\otimes_M m=r\otimes_M(s\otimes_M m)\\
    \textit{commutativity:}
      & m\oplus_M n=n\oplus_M m &\\
    \textit{distributivity:}
      & r\otimes_M(m\oplus_M  n)=(r\otimes_M m)\oplus_M (r\otimes_M n)
      & (r\algplus s)\otimes_M m=(r\otimes_M m)\oplus_M (s\otimes_M m)\\
    \textit{annihilation:}
      & \algzero\otimes_M m=\boldsymbol{0}_M
      & r\otimes_M\boldsymbol{0}_M=\boldsymbol{0}_M
  \end{array}
\]
When $\tau$ is commutative, we call $M$ a \emph{commutative semimodule}.

\subsection{Syntax}

Fix a commutative semiring $\tau$.
The syntax of kernels and signals is shown in \autoref{fig:syntax}.
For the purpose of this section,
 kernels and signals are $d$-dimensional.
\autoref{sec:compiler} specializes to 1D and 2D kernels and signals and discusses their efficient code generation.
Throughout the paper we use
$F, G, H$ as metavariables for signals
and
$K, A, B, C$ for kernels.
We often denote by $G$ the external input signal
and by $F$ the computed output signal.

A signal $F$ is either an input variable $x$, a constant signal $c$ for $c\in \tau$ (with slightly abused notation),
an impulse $\delta_c$, 
a pointwise addition $F\oplus G$, a pointwise multiplication $F\odot G$,
a convolution $K\otimes F$,
a recurrence $\rec(K, G)$, 
or $\ominus F$ if $\tau$ is a ring.
Note that convolutions and recurrences are defined between kernels and signals.

\begin{figure}[ht]
\begin{minipage}[t]{0.48\textwidth}
\begin{center}
\small
\setlength{\arraycolsep}{4pt}
\[
\begin{array}{lrcl}
    \text{Signal}
      & F,G,H \in \mathcal{S}& ::= & x \ \mid\ c \ \mid\ \delta_c \ \mid\  K[\veci] \\
      & & \mid & F \oplus G \ \mid\ F\odot G\\
      & & \mid & K\otimes F\\
      & & \mid &  \rec(K,\; G) \\
      & & \mid & \ominus F  \quad \text{if $\tau$ is a ring}\\[6pt]
    \text{Kernel}
    & K,A,B,C \in \mathcal{K}&::= & [F_\veci]_{\veczero\leq \veci\leq \vecn}  \\
      &&\mid&  K_1 \oplus K_2\ \mid\ K_1\otimes K_2 \\
      &&\mid& \ominus K \quad \text{if $\tau$ is a ring}\\[6pt]
    \end{array}
    \]
\end{center}
\captionof{figure}{Syntax for signals and kernels.}
\label{fig:syntax}
\end{minipage}
\hfill
\begin{minipage}[t]{0.48\textwidth}
\begin{center}
\small
\[
\begin{aligned}
\order\big([F_\veci]_{\veczero \leq \veci \leq \vecn}\big)
  &= \vecn \\
\order(K_1 \oplus K_2)
  &= \max\left(\order(K_1),\order(K_2)\right) \\
\order(K_1 \otimes K_2)
  &= \order(K_1) + \order(K_2) \\
\order(\ominus K)
  &= \order(K)
\end{aligned}
\]
\end{center}
\captionof{figure}{The order defines the domain of a kernel.
In the context of kernel orders, $\max{}$ is interpreted as the dimension-wise maximum of two vectors.
Similarly, we say $\veci\leq \vecn$ if $\veci$ is less than or equal to $\vecn$ in every dimension.}
\label{fig:kernel-order}
\end{minipage}
\end{figure}


A kernel is a finite $d$-dimensional array of signals 
 $[F_\veci]_{\veczero\leq \veci\leq \vecn}$.
We call the signals of a kernel its \emph{taps}.
A kernel is called a singleton if it only has one tap.
Kernels can be tap-wise added
 ($K_1\oplus K_2$)
 or convolved ($K_1\otimes K_2$).
A kernel can be additively inverted ($\ominus K$) if $\tau$ is a ring.
We call $[\algone]$ the identity kernel $I$ and
 $[\algzero]$ the zero kernel $\zerokernel$.
The \emph{order} defines the domain of a kernel and is shown in \autoref{fig:kernel-order}.
We define the time-delay kernel $z^{-\veck}$ so that
 $\left(z^{-\veck}\right)_\veck=\algone$ and is $\algzero$ everywhere else.
For instance, the one-dimensional delay kernel $z^{\compactsub 1}$ is defined as $[\algzero, \algone]$,
 and in the degenerate case $z^{\veczero}=I$.
We use $A^k$ to denote $A$ convolved with itself $k$ times, $A\otimes A\otimes \ldots\otimes A$,
 and $A^0=I$. This notation will be justified later as we show convolution is associative (\autoref{sec:conv-algebra}).

\subsection{Semantics}
\label{sec:semantics}

\begin{figure}[t]
\footnotesize
\begin{align*}
\llbracket x \rrbracket^\sigma &= \lambda \vecn. \texttt{if}\ \vecn\geq \veczero\ \texttt{then}\ \sigma[x](\vecn) \ \texttt{else}\ \algzero\\
\llbracket c \rrbracket^\sigma &= \lambda \vecn. \texttt{if}\ \vecn\geq \veczero\ \texttt{then}\ c \ \texttt{else}\ \algzero\\
\llbracket \delta_{c} \rrbracket^\sigma &= \lambda \vecn. \texttt{if}\ \vecn=\veczero\ \texttt{then}\ c \ \texttt{else}\ \algzero\\
\llbracket K[\veci]\rrbracket^\sigma&=\begin{cases}
    \ \left \llbracket K\right\rrbracket^\sigma_\veci &\text{if $\veczero\leq \veci\leq\order(K)$}\\
    \ \left\llbracket\algzero\right\rrbracket^\sigma & \text{otherwise}
\end{cases}\\
\llbracket F\oplus G\rrbracket^\sigma &=\lambda \vecn. \llbracket F\rrbracket^\sigma (\vecn)\algplus \llbracket G\rrbracket^\sigma(\vecn)\\
\llbracket F\odot G\rrbracket^\sigma &=\lambda \vecn. \llbracket F\rrbracket^\sigma (\vecn)\algtimes \llbracket G\rrbracket^\sigma (\vecn)\\
\llbracket K\otimes F\rrbracket^\sigma &
      =\lambda \vecn. \bigoplus_{\veczero\leq \veci\leq\order(K)}\llbracket K\rrbracket^\sigma_i(\vecn)\algtimes \llbracket F\rrbracket^\sigma (\vecn-\veci)
      =\semof{\bigoplus_{\veczero\leq\veci\leq\order(K)} K[\veci]\odot \left(\zk{-\veci}\otimes  F\right)}
    \\
\llbracket \rec(K,G)\rrbracket^\sigma &= \lambda \vecn. \bigoplus_{k\geq 0} \llbracket K^k\otimes G\rrbracket^\sigma (\vecn) = \semof{\bigoplus_{k=0}^\infty K^k\otimes G} \\
\llbracket \ominus F \rrbracket^\sigma &=\lambda \vecn. \algminus \llbracket F\rrbracket^\sigma (\vecn)\\
\llbracket [F_\veck]_{\veczero\leq \veck\leq \vecn}\rrbracket^\sigma&=\left[\left\llbracket F_\veck\right\rrbracket^\sigma\right]_{\veczero\leq \veck \leq \vecn}\\
\llbracket K_1\oplus K_2\rrbracket^\sigma&=\Big( \left\llbracket K_1[\veck]\oplus  K_2[\veck]\right\rrbracket^\sigma   \Big)_{\veczero\leq \veck\leq \order(K_1\oplus K_2) }\\
\llbracket K_1\otimes K_2\rrbracket^\sigma 
                    &= \Bigg[
                        \lambda \vecn. \bigoplus_{\veczero\leq \veci\leq\veck}\semof{K_1}_{\veci}(\vecn)\algtimes \semof{K_2}_{\veck-\veci}(\vecn-\veci)
                        \Bigg]_{\veczero\leq \veck\leq \order(K_1\otimes K_2)}\\
                    &=\Bigg[
                        \left\llbracket \bigoplus_{\veczero\leq \veci\leq\veck} K_1[\veci]\odot \left(z^{-\veci}\otimes K_2[\veck-\veci]\right) \right\rrbracket^\sigma
                    \Bigg]_{\veczero\leq \veck\leq \order(K_1\otimes K_2)}\\
\llbracket \ominus K\rrbracket^\sigma &=  \Big[ \left\llbracket\ominus K[\veck] \right\rrbracket^\sigma\Big]_{\veczero\leq \veck\leq\order(K)}
\\
\end{align*}
\caption{Semantics for signals and kernels.}
\label{fig:1d-semantics}
\end{figure}

\rev{The semantics of a signal over semiring $\tau$ is a mathematical function $\mathbb{Z}^d\to\tau$
that maps each time index $\vecn$ to a value in $\tau$,
 and the semantics of a kernel is a finite, multi-dimensional array of signals, indexable by an offset $\veci\in\mathbb{Z}^d$.
We say the signals in the array the \emph{taps} of the kernel.}
\rev{
  Fix an environment $\sigma$ that maps variables to semantic signal $\mathbb{Z}^d\to\tau$.
  We write $\semof{F}$ and $\semof{K}$ for the semantics of signal $F$ and kernel $K$, respectively.
  \autoref{fig:1d-semantics} shows the definitions of $\semof{F}$ and $\semof{K}$.
}

We say a signal $F$ is \emph{time-invariant} if its semantics is a constant function, i.e., $\forall \vecn_1,\vecn_2. \semof{F}(\vecn_1)=\semof{F}(\vecn_2)$.
We say a kernel $K$ is \emph{time-invariant} if all of its taps are time-invariant.
Otherwise the kernel is called \emph{time-varying}.

\paragraph*{Signals}
Signals evaluate to functions from time indices to values in $\tau$.
They evaluate to $\algzero$ at any negative index.
A signal variable is evaluated by looking up the environment $\sigma$, 
 and a constant signal $c$ is evaluated to $c$ at every non-negative index.
The impulse $\delta_c$ is $c$ at the origin and $\algzero$ elsewhere.
Tap selection $K[\veci]$ returns the taps of kernel $K$ at offset $\veci$ if
 defined and the zero signal $\algzero$ otherwise.
The operators $F\oplus G$ and $F\odot G$ lift $\algplus$ and $\algtimes$
 to signals pointwise.

The expression $A\otimes F$
  convolves kernel $A$ with signal $F$:
  it slides $A$ across $F$, multiplying each tap by
  the correspondingly shifted signal and summing.
At time $\vecn$,
\[
  \semof{A\otimes F}(\vecn)
  = \bigoplus_{\veczero\leq \veci\leq\order(A)}
      \semof{A}_\veci(\vecn)\algtimes\semof{F}(\vecn-\veci).
\]


Convolving with the delay kernel $z^{-\veci}$ delays the signal by $\veci$ time steps, so
 $z^{-\veci}\otimes F$ is $F$ shifted forward in time by $\veci$.
In fact, any convolution $A\otimes F$ can be written in terms of convolution with the delay kernel, i.e.,
 $\bigoplus_{\veci} A[\veci]\odot \left(\zk{-\veci}\otimes F\right)$.

\begin{example}
A finite impulse response (FIR) filter in signal processing is a convolution $A\otimes G$ between a time-invariant kernel $A$ 
 and a signal $G$
 over the arithmetic ring.
Consider the moving sum $f(x) = g(x) + g(x\compactsub 1) + g(x\compactsub 2) + g(x\compactsub 3)$.
This is represented in the recurrence algebra as $F = [1,\ 1,\ 1,\ 1]\otimes G$.
\end{example}

The recurrence $\rec(A,G)$ denotes the unbounded sum $\bigoplus_{k\geq 0} A^k\otimes G$.
Equivalently, it computes the signal $F$ satisfying $F=A\otimes F\oplus G$.
The infinite sum is well-defined when $\semof{A}$ is $\algzero$ at offset $\veczero$.
We call this condition \emph{causality}, and it ensures that the recurrence does not depend on its own current value.
We defer a detailed discussion of $\rec$ operator to \autoref{sec:algebra}.

\begin{example}
An infinite impulse response (IIR) filter in signal processing is a recurrence $\rec(K, G)$
 between a time-invariant kernel $K$ and an input signal $G$ over the arithmetic ring.
Our band-pass IIR filter \(f(x) = 1.8 \cdot f(x\compactsub1) - 0.9 \cdot f(x\compactsub2) + g(x)\) is represented in the recurrence algebra
  as $F = \rec([0, 1.8, -0.9], G)$, which is a second-order, time-invariant recurrence.
\end{example}

\paragraph*{Kernels}
A kernel denotes an array of signals:
 a literal $[F_\veci]_{\veczero \leq \veci \leq \vecn}$ denotes the array of denotations of its taps $F_\veci$,
 $K_1\oplus K_2$ adds the kernels tap-wise (widening to the higher order),
 and $\ominus K$ negates $K$ tap-wise.

$K_1\otimes K_2$ convolves the two kernels.
The $\veck$-th tap of the convolution collects every way of splitting $\veck$ across the two operands;
 it pairs a tap $\semof{K_1}_{\veci}$ at delay $\veci$ with a tap $\semof{K_2}_{\veck-\veci}$ at delay $\veck-\veci$ for $\veczero\leq \veci\leq \veck$.
This splitting also shifts when the second coefficient is read.
 Recall that a tap at index $\veci$ denotes a contribution from $\veci$ steps back.
 Since the tap of $K_1$ already sits $\veci$ steps in the past,
 the coefficient of $K_2$ that meets it is taken from $\veci$ steps earlier as well, hence $\semof{K_1}_{\veci}(\vecn)\algtimes \semof{K_2}_{\veck-\veci}(\vecn-\veci)$.
If $K_2$ is time-invariant, its coefficients do not depend on time,
 so the delay has no effect
 and the expression recovers the familiar convolution sum.
Putting everything together, we have the convolution evaluates to $\lambda \vecn.\bigoplus_{\veczero\leq \veci\leq \veck} \semof{K_1}_{\veci}(\vecn)\algtimes \semof{K_2}_{\veck-\veci}(\vecn-\veci)$ at tap $\veck$.
Equivalently, we can also use the delay kernel to denote the time shift and denote the tap $\veck$ of the convolution as $\left\llbracket \bigoplus_{\veczero\leq \veci\leq\veck} K_1[\veci]\odot \left(z^{-\veci}\otimes K_2[\veck-\veci]\right) \right\rrbracket^\sigma$.
This form is handy for algebraic simplification.

\begin{example}
  Consider time-varying kernels $A = [S_1, S_2]$ and $B = [S_3, \algzero, S_4]$.
  The convolution $A\otimes B$ is equivalent to
    \begin{align*}
    A\otimes B &= [\,
        S_1\odot S_3, \ \ 
        S_1\odot \algzero\oplus S_2\odot (\zk{\compactsub 1}\otimes S_3),\ \ 
        S_1\odot S_4\oplus S_2\odot (\zk{\compactsub 1}\otimes \algzero), \ \ 
        S_2\odot (\zk{\compactsub 1}\otimes S_4)\,
    ]\\
    &= [\,
        S_1\odot S_3, \ \ 
        S_2\odot (\zk{\compactsub 1}\otimes S_3),\ \ 
        S_1\odot S_4, \ \ 
        S_2\odot (\zk{\compactsub 1}\otimes S_4)\,
    ]
    \tag*{\qedhere}
    \end{align*}
\end{example}




\subsection{Properties of $\oplus$ and $\otimes$.}
\label{sec:conv-algebra}

From the semantics, we can derive the algebraic properties of $\oplus$ and $\otimes$.

\begin{proposition}
The operators $\oplus$ and $\otimes$ have the following properties:
\[
  \small
  \renewcommand{\arraystretch}{1.2}
  \begin{array}{r@{\quad}l@{\qquad}l}
    \textit{associativity:}
      & (A\otimes B)\otimes C = A\otimes (B\otimes C)
      & (A\otimes B)\otimes G = A\otimes (B\otimes G)\\
    \textit{distributivity:}
      & (A\oplus B)\otimes G = A\otimes G\oplus B\otimes G
      & A\otimes (G\oplus H) = A\otimes G\oplus A\otimes H
  \end{array}
\]
If $A$ and $B$ are time-invariant,
  we additionally have commutativity
    \(A\otimes B=B\otimes A.\)
\end{proposition}

Convolution is associative across both kernel-kernel and kernel-signal products,
 and it distributes over addition on both sides.
Convolution is in general not commutative,
 because the order in which a tap is applied determines the time index at which its coefficient is read;
 when both kernels are time-invariant, however,
 convolution becomes commutative.

In fact, kernels form a semiring, and the time-invariant kernels form a commutative sub-semiring.
\begin{proposition}
    Let $\mathcal{K}$ be the set of all kernels.
    $(\mathcal{K}, \oplus, \otimes, \zerokernel, I)$ is a semiring.
    Additionally, let $\mathcal{K}_\textit{TI}$ be the set of time-invariant kernels.
    Then
    $(\mathcal{K}_\textit{TI}, \oplus, \otimes, \zerokernel, I)$ is a commutative semiring.
\end{proposition}

Kernel-signal convolution connects kernels with signals,
 and it is naturally compatible with the kernel semiring. Indeed, the signals form a semimodule over the kernel semiring.

\begin{proposition}
    Let $\mathcal{S}$ be the set of all signals.
    $(\mathcal{S}, \algzero, \oplus, \otimes)$ forms a semimodule over the semiring $(\mathcal{K}, \oplus, \otimes, 0, I)$, and it forms a commutative semimodule over the semiring $(\mathcal{K}_{\textit{TI}}, \oplus, \otimes, \zerokernel, I)$.
\end{proposition}

To summarize, semirings and semimodules capture the algebraic structure of kernels and signals.
The following example demonstrates the algebraic properties.

\begin{example}[Convolving with time-invariant kernels]
Consider again the 4-tap moving sum $F = K\otimes G$ with $K = [1,1,1,1]$.
We can factor the kernel $K$ into $A = [1,0,1]$ and $B = [1,1]$,
 and $K=A\otimes B$.
By associativity, we have
 $F =K\otimes G= (A\otimes B)\otimes G = A\otimes(B\otimes G)$.
Additionally, because $A$ and $B$ are time-invariant, we also have
 $F=(B\otimes A)\otimes G=B\otimes(A\otimes G)$ by commutativity.
They correspond to different implementations of the moving sum.
For instance, $A\otimes (B\otimes G)$ corresponds to a cascade of two filters:
 the inner filter $B\otimes G$ first computes the pointwise sums $h(x) = g(x)+g(x\compactsub 1)$,
 and the outer filter $A$ then adds pairs of $h$ that are two elements apart,
 i.e., $f(x) = h(x) + h(x\compactsub 2)$.
\end{example}


A one-dimensional time-invariant kernel corresponds to a polynomial:
$A=[a_0,\ldots,a_n]$ is the coefficient list of
$p_A(x)=\sum_i a_i x^i$, and convolution of time-invariant kernels
is multiplication of their polynomials.
By the fundamental theorem of algebra,
  any real polynomial factors into real factors of degree at most two,
  which bounds the order
 of a kernel's factors.
 


\begin{lemma}
    Under the arithmetic ring over real numbers,
    any one-dimensional time-invariant kernel $A$
    of order $n$ can be factored as
    $A = A_1\otimes \ldots\otimes A_k$ where each $A_i$ 
    has order less than or equal to 2.
\end{lemma}

See the full version for the proof.
Signal processing often uses this factoring
  to trade off numerical stability and performance in filter design.
For example, the moving-sum kernel
  $[1,1,1,1]$ is the polynomial
  $1+x+x^2+x^3=(1+x^2)(1+x)$,
  which is exactly the factorization $[1,0,1]\otimes[1,1]$ above.

The following example shows why convolutions in general do not commute.

\begin{example}[Convolving with time-varying kernels]
    Consider again time-varying kernels $A = [S_1, S_2]$ and $B = [S_3, \algzero, S_4]$.
  Let $F = A \otimes (B \otimes G)$.
  By associativity,
    $F = (A\otimes B)\otimes G$.
    Meanwhile, $(B\otimes A)\otimes G$
    computes a different signal.
    \begin{align*}
    B\otimes A
    &= [\,
        S_3\odot S_1, \ \
        S_3\odot S_2, \ \
        S_4\odot (\zk{-2}\otimes S_1), \ \
        S_4\odot (\zk{-2}\otimes S_2)\,
    ]
    \neq A\otimes B
    \end{align*}
    To confirm this in the explicit indexed form, let $s_1,s_2,s_3,s_4$ denote $\semof{S_1},\semof{S_2},\semof{S_3},\semof{S_4}$.
{\footnotesize
\begin{equation*}
    \begin{aligned}
    h(n)  &= s_3(n)\,g(n) + s_4(n)\,g(n\compactsub 2)\\
    f(n)  &= s_1(n)\,h(n)  + s_2(n)\,h(n\compactsub 1)\\
    &= 
    \begin{aligned}[t]
    &s_1(n)\,s_3(n)\,\cdot\, g(n)
    \;+\; s_2(n)\,s_3(n\compactsub 1)\,\cdot\,g(n\compactsub 1)\\
    \;+\; &s_1(n)\,s_4(n)\,\cdot\,g(n\compactsub 2)
    \;+\; s_2(n)\,s_4(n\compactsub 1)\,\cdot\,g(n\compactsub 3)
    \end{aligned}
    \end{aligned}
    \begin{aligned}
    h'(n) &= s_1(n)\,g(n) + s_2(n)\,g(n\compactsub 1)\\
    f'(n) &= s_3(n)\,h'(n) + s_4(n)\,h'(n\compactsub 2)\\
    &=
    \begin{aligned}[t]
    &s_3(n)\,s_1(n)\,\cdot\, g(n)
    \;+\; s_3(n)\,s_2(n)\,\cdot\,g(n\compactsub 1)\\
    \;+\; &s_4(n)\,s_1(n\compactsub 2)\,\cdot\,g(n\compactsub 2)\\
    \;+\; &s_4(n)\,s_2(n\compactsub 2)\,\cdot\,g(n\compactsub 3).
    \end{aligned}
    \end{aligned}
\end{equation*}
}
    
    The functions $f$ and $f'$ match the definitions of 
    $F=(A\otimes B)\otimes G$ and $F'=(B\otimes A)\otimes G$,
    and they are not equivalent.
\end{example}

To summarize, this section presents the syntax and semantics of the recurrence algebra together with the properties of $\otimes$ and $\oplus$.
Next, we turn to the recurrence operator.

\section{The Recurrence Operator}
\label{sec:algebra}

The recurrence operator \rec{} reifies the fixed point computation of recurrences
 into the language of kernels and signals.
We first show that the semantics of $\rec(A, G)$ is well defined
 and is equivalent to finding an $F$ that satisfies $F=A\otimes F\oplus G$.

\begin{lemma}[Equivalent definitions of \rec{}]
    Suppose $A$ is causal.
    $\bigoplus_{i\geq 0} A^i\otimes G$ is well-defined, and 
    $F = \bigoplus_{i\geq 0} A^i\otimes G$ iff $F=A\otimes F\oplus G$.
\end{lemma}

In the rest of the section,
 we present the algebraic identities of the \rec{} operator.
We start with identities that hold for any semiring and then specialize to rings and idempotent semirings.
Some of these identities directly enable vectorization,
 while others capture and give insight into transformations that practitioners use.
\autoref{fig:identities-summary} summarizes these identities.
We comment on their contexts and implications when applicable.
The proofs are algebraic and deferred
 to the appendix of the full version.

\begin{figure}[t]
\centering
\fbox{\begin{minipage}{0.93\textwidth}

\emph{\textbf{Any commutative semiring}}
{\footnotesize
\begin{align*}
\rec(A, G) &= \rec\bigl(A^k,\ \textstyle\bigoplus_{i=0}^{k-1} A^i\otimes G\bigr)
  \tag*{\textit{Unrolling}}\\
\rec(A, B\otimes G) &= B\otimes \rec(A, G) \hspace{2em}(A,B\text{ t.i.})
  \tag*{\textit{Commuting recurrence with convolution}}\\
\rec(A, \rec(B, G)) &= \rec(B, \rec(A, G)) \hspace{2em}(A,B\text{ t.i.})
  \tag*{\textit{Commuting recurrences}}
\end{align*}
}

\emph{\textbf{Ring (e.g., arithmetic ring)}}
{\footnotesize
\begin{align*}
(I\ominus A)\otimes \rec(A,G) &= G
  \tag*{\textit{Inverse of Recurrence}}\\
\rec(A, \rec(B, G)) &= \rec(A\oplus B\ominus B\otimes A,\ G)
  \tag*{\textit{Fusion of Stacked Ring Recurrences}}\\
\rec(A, G)\oplus\rec(B, H) &= \rec\bigl(A\oplus B\ominus A\otimes B,\ (I\ominus B)\otimes G\oplus(I\ominus A)\otimes H\bigr)
  \hspace{1em}(A,B\text{ t.i.})
  \tag*{\textit{Fusion of Parallel Ring Recurrences}}\\
\rec(A\oplus B,G) &= \rec\bigl(A\oplus A\ominus A^2\oplus B^2,\ (I\ominus A\oplus B)\otimes G\bigr)
  \hspace{1em}(A,B\text{ t.i.})
  \tag*{\textit{Dilation with Additive Inverse}}\\
\rec(A^2, G) &= \rec(A, \rec(\ominus A, G))
  \tag*{\textit{Decomposition of Squared Kernels}}\\
\rec(A, G)&= \rec(A\ominus B\otimes(I\ominus A),\ (I\oplus B)\otimes G)
  \tag*{\textit{Change of Basis}}
\end{align*}
}

\emph{\textbf{Idempotent semiring (e.g., tropical semiring)}}
{\footnotesize
\begin{align*}
\rec(A\oplus B, G) &= \rec(A\oplus B^2,\ (I\oplus B)\otimes G) \hspace{1em}(A,B\text{ t.i.})
  \tag*{\textit{Dilation with Idempotence}}\\
\rec(A, \rec(B, G)) &= \rec(A\oplus B, G) \hspace{1em}(A,B\text{ t.i.})
  \tag*{\textit{Fusion of Idempotent Recurrences}}
\end{align*}
}

\end{minipage}}
\caption{Recurrence identities grouped by the class of semirings.
``t.i.''\ abbreviates time-invariant.}
\label{fig:identities-summary}
\end{figure}

\subsection{Identities for Recurrences over Any Semiring}
\label{sec:identities}

We start with identities that hold over any semiring.
First, we can unroll a recurrence.

\begin{lemma}[Unrolling]
  For kernel $A$ and signal $G$ over any semiring,
\label{lem:unrolling}
\[
    \rec(A, G) = \rec\bigl(A^k,\ \textstyle\bigoplus_{i=0}^{k-1} A^i\otimes G\bigr).
\]
\end{lemma}

Intuitively, instead of advancing the recurrence one step at a time with $A$, we
may take $k$ steps at once using the unrolled kernel $A^k$, after pre-accumulating
the contributions of the $k$ intermediate steps, $\bigoplus_{i=0}^{k-1} A^i\otimes G$,
into the input.

The next two lemmas concern the commutativity of recurrences.
Recall that time-invariant convolutions commute with each other.
Indeed, time-invariant convolutions also commute with time-invariant recurrences, and
 time-invariant recurrences commute with each other.

\begin{lemma}[Commutativity of recurrences]
\label{lem:kernel-factor}
\label{lem:nested-comm}
For time-invariant kernels $A$ and $B$ and signal $G$ over any semiring,
{\small
\[
    \rec(A, B\otimes G) = B\otimes \rec(A, G).
\]
\[
    \rec(A, \rec(B, G)) = \rec(B, \rec(A, G)).
\]
}
\end{lemma}

These identities are helpful when implementing cascades of convolutions and recurrences.

\begin{example}
  In image processing, repeated applications of box filters~\citep{recursive-gaussian}~are often used to approximate Gaussian filters.
  A box filter can be efficiently implemented as a cascade of a recurrence and a convolution: $\textsf{Box}(G) = B\otimes \rec(A, G)$,
  where the 2D kernel $A$ computes the summed area table of the input image,
  and the 2D kernel $B$ computes the difference of the summed area table.
  Therefore, a Gaussian filter can be approximated as $\textsf{Gauss}(G) = B\otimes \rec(A, B\otimes \rec(A, B\otimes \rec(A, G)))$.
  Using commutativity, we can freely reorder the recurrences and convolutions in the cascade.
  For instance, we can fuse the convolutions together and derive $\textsf{Gauss}(G) = B\otimes B\otimes B\otimes \rec(A, \rec(A, \rec(A, G)))$.
\end{example}

\subsection{Identities for Recurrences over a Ring}

This section presents the identities that hold for recurrences over a ring.

\begin{lemma}[Inverse of Recurrence]
\label{lem:telescoping}
For kernel $A$ and signal $G$ over a ring,
\[
    (I\ominus A)\otimes \rec(A,G) = G.
\]
\end{lemma}

The above lemma shows that when restricted to rings,
 convolution by $I\ominus A$ is precisely the inverse to the recurrence $\rec(A,\cdot)$.
As an analogy, we can view $\rec(A, G)$ as algebraically similar to computing the geometric series $\sum_{i\geq 0 }ar^i$,
 and it holds that $(1-r)\cdot\sum_{i\geq 0} ar^i = a$ when the series converges.
Similarly, applying $I\ominus A$ undoes the
 recurrence and recovers the original input.

The next identity flattens a stack of ring recurrences into a single recurrence with a larger kernel.

\begin{lemma}[Fusion of Stacked Ring Recurrences]
\label{lem:flatten}
For kernels $A$ and $B$ and signal $G$ over a ring,
\[
    \rec(A, \rec(B, G)) = \rec(A\oplus B\ominus B\otimes A,\,G).
\]
\end{lemma}

Sometimes we apply this identity in the opposite direction,
 decomposing a single recurrence into a stack of recurrences with smaller kernels.

 \begin{example}
  In the repeated box filter example, the recurrence kernel $A=\left[\begin{smallmatrix}
    0, 1\\
    1, -1
  \end{smallmatrix}\right]$ can be decomposed into 
  a horizontal kernel $A_x=[0, 1]$ and a vertical kernel $A_y=\left[\begin{smallmatrix}0\\1\end{smallmatrix}\right]$ using \autoref{lem:flatten}.
  Therefore, the implementation space of the repeated box filter
  consists of permutations of repeated convolutions with $B$ and repeated recurrences with $A_x$ and $A_y$.
\end{example}

As a small digression, by the fundamental theorem of algebra, we can always decompose a one-dimensional time-invariant recurrence over the reals.
\begin{lemma}
  Under the arithmetic ring over real numbers,
  the recurrence $\rec(A, G)$ for which $A$ is a 1D time-invariant kernel of order $n$ 
  can be decomposed into a stack of recurrences $\rec(A_1, \allowbreak\rec(A_2, \allowbreak\ldots \rec(A_k, G)))$ 
  where each $A_i$ has order at most 2.
\end{lemma}

For time-invariant kernels,
 we have a similar identity for fusing a \emph{parallel} sum of recurrences.

\begin{lemma}[Fusion of Parallel Ring Recurrences]
\label{lem:parallel-merge}
For time-invariant kernels $A$ and $B$ and signals $G$ and $H$ over a ring,
\[
    \rec(A, G)\oplus\rec(B, H) = \rec\bigl(A\oplus B\ominus A\otimes B,\,(I\ominus B)\otimes G\oplus(I\ominus A)\otimes H\bigr).
\]
\end{lemma}

Applying the above lemma in the opposite direction
 decomposes a time-invariant recurrence into a parallel sum of recurrences.
This is known as partial fraction expansion.
Partial fraction expansion can help parallelize a high-order recurrence
 by decomposing it into simpler independent recurrences.
Practitioners use the fusion and decomposition of stacked and parallel recurrences
 to trade off performance, parallelism, and numerical stability in digital filter design.

The next identity is the foundation of dilation for time-invariant ring recurrences.

\begin{lemma}[Dilation with Additive Inverse]
\label{lem:ring-dilation}
For time-invariant kernels $A$ and $B$ and signal $G$ over a ring,
\[
    \rec(A\oplus B,G) = \rec\bigl(A\oplus A\ominus A^2\oplus B^2,\,(I\ominus A\oplus B)\otimes G\bigr).
\]
\end{lemma}

This identity proves the transformation in the motivating example in \autoref{sec:overview-example}
 by treating $A$ and $B$ as the even and odd splits of the original kernel.

The next two identities show how to decompose a recurrence with $A^2$ and how to change the kernel used in a recurrence;
 both follow from the identities above.

\begin{lemma}[Decomposition of Squared Kernels]
\label{lem:square-split}
For kernel $A$ and signal $G$ over a ring,
\[
    \rec(A^2, G) = \rec(A, \rec(\ominus A, G)).
\]
\end{lemma}

\begin{lemma}[Change of Basis]
\label{lem:basis-change}
For kernels $A$ and $B$ and signal $G$ over a ring,
\[
    \rec(A, G) = \rec(A\ominus B\otimes(I\ominus A),\ (I\oplus B)\otimes G)
\]
\end{lemma}

\subsection{Identities for Recurrences over an Idempotent Semiring}

The idempotence of addition, $a\algplus a=a$,
 enables a different set of identities.
The identity below dilates recurrences over idempotent semirings.

\begin{lemma}[Dilation with Idempotence]
\label{lem:idem-dilation}
For time-invariant kernels $A$ and $B$ and signal $G$ over an idempotent semiring,
\[
    \rec(A\oplus B, G) = \rec(A\oplus B^2,\ (I\oplus B)\otimes G).
\]
\end{lemma}

This identity allows us to break a kernel into a sum of sub-kernels and dilate them independently.

\begin{example}
\label{exm:dist-trans}
  Consider distance transform $F = \rec\left(\left[\begin{smallmatrix}
    \infty, 1\\
    1,\sqrt{2}
\end{smallmatrix}\right], G\right)$ over the $(\min,+)$-semiring.
Let $A = \left[\begin{smallmatrix}
    \infty, \infty\\
    1,\sqrt{2}
\end{smallmatrix}\right]$ and $B = \left[\begin{smallmatrix}
    \infty, 1\\
    \infty,\infty
\end{smallmatrix}\right]$, the dilation identity gives us $F=\rec\left(\left[\begin{smallmatrix}
    \infty, \infty,2\\
    1,\sqrt{2},\infty
\end{smallmatrix}\right], \left[\begin{smallmatrix}
  0, 1\\
  \infty, \infty
\end{smallmatrix}\right]\otimes G\right)$.
Applying the identity again by setting $B$ to be the two other taps of the original kernel, we get
\[
F=\rec\left(\left[\begin{smallmatrix}
    \infty, &\infty,&2\\
    \infty,&\infty,&\infty\\
    2,&\infty,&2\sqrt{2}
\end{smallmatrix}\right], \left[\begin{smallmatrix}
  0, &\infty\\
  \infty, &\sqrt{2}
\end{smallmatrix}\right]
\otimes
\left[\begin{smallmatrix}
  0, &\infty\\
  1, &\infty
\end{smallmatrix}\right]
\otimes 
\left[\begin{smallmatrix}
  0, &1\\
  \infty, &\infty
\end{smallmatrix}\right]\otimes G\right).
\]
This computes two consecutive elements (i.e., $f(x, y)$ and $f(x, y+1)$) at a time.
Moreover, because the new $f$ no longer depends on itself from the previous row,
 two rows can be computed in parallel with multi-cores.
By choosing which tap to dilate, we  obtain the desired shape of parallelism.
For instance, dilating the tap at the zeroth row again computes four horizontal elements at a time with vectorization,
 while leaving the multi-core parallelism unchanged.
\end{example}

We can also fuse stacked recurrences
 by simply summing the kernels in the idempotent semirings.

\begin{lemma}[Fusion of Idempotent Recurrences]
\label{lem:parallel-decomp}
For time-invariant kernels $A$ and $B$ and signal $G$ over an idempotent semiring,
\[
    \rec(A, \rec(B, G)) = \rec(A\oplus B, G)
\]
\end{lemma}

Compare this with fusion of stacked recurrence over a ring.
In general, 
 a stack of recurrences can double count the contribution of an algebraic path.
For rings, this is corrected using subtraction,
 whereas for idempotent rings, double counting does not change the result.

\section{Techniques for Vectorizing Recurrences}
\label{sec:vectorization-tricks}

We now present techniques for vectorizing different types of recurrences.
This section focuses on 1D recurrences and defers higher-dimensional recurrences to \autoref{sec:extension}.
We present four techniques: dilation for ring recurrences, dilation for idempotent recurrences,
 fast associative scan, and path-based decomposition.
The first two follow from the algebraic identities of the previous section,
 and they apply only when the kernel is time-invariant and the semiring satisfies certain properties.
Fast associative scan builds on a general transformation for prefix scans over associative operations~\citep{kogge-stone,BlellochTR90}.
Path-based decomposition views convolution as aggregating over algebraic paths,
 and decomposes those paths by divide-and-conquer to expose vectorizable work.
Unlike dilations, the latter two techniques apply to any type of recurrence,
 at the cost of smaller speedup.

Some of these techniques are not new. 
They have appeared in various forms in the literature,
 for example in multi-core parallelism or digital signal processing,
 usually within a specific domain.
In relation to these works,, our contribution is twofold: (1) we show that these techniques
 apply to SIMD vectorization, and (2) we distill them into an algebraic framework
 and, in some cases, generalize them.
\autoref{sec:related} discusses related work and our extensions to it.

\subsection{Dilations for Time-Invariant Recurrences}
\label{sec:dilation}

We present two dilation techniques, one for recurrences over rings and the other over idempotent semirings.
Both require the recurrence to be time-invariant,
 because the underlying identities rely on commutativity.
We call these techniques dilation because both transform a recurrence into another that
 doubles the \emph{strides} between, or ``dilates'', the taps of the kernel.
Formally, we define the stride $s$ of a 1D kernel $K$ as the largest integer such that $K[i]$ evaluates to the constant signal $\algzero$ for all $i$  not divisible by $s$.
For instance, the stride of kernel $[0, a, b]$ is 1, and the stride of kernel $[0, 0, a, 0, b]$ is 2.

\paragraph*{Ring recurrences}

We use the identity in \autoref{lem:ring-dilation} to dilate ring recurrences.
Recall that 
\(\rec(A\oplus B,G) = \rec\bigl(A\oplus A\ominus A^2\oplus B^2,\,(I\ominus A\oplus B)\otimes G\bigr)\)
 for time-invariant kernels $A$ and $B$.
Let $K$ be a kernel with stride $s$ and $F=\rec(K, G)$.
Define $A$ as the kernel that only contains the taps of $K$ at multiples of $2s$,
 and $B$ as the kernel that only contains the taps of $K$ at indices that are $s$ modulo $2s$.
We have $K=A\oplus B$.
Applying the dilation identity 
 yields $F=\rec\bigl(A\oplus A\ominus A^2\oplus B^2,\,(I\ominus A\oplus B)\otimes G\bigr)$.
We claim that the new kernel $A\oplus A\ominus A^2\oplus B^2$ has a stride of $2\cdot s$.
To see this, first, $A$ has a stride of $2\cdot s$,
 and so does $A\oplus A\ominus A^2$.
Second, suppose $B^2$ has a non-zero tap at index $k$.
There must exist $k_1,k_2$ with $k_1+k_2=k$ such that $B[k_1],B[k_2]$ are non-zero.
As $k_1\mod 2s=s$ and $k_2\mod 2s=s$, it follows that $(k_1+k_2) \mod 2s = 0$, so $B^2$ also has a stride of $2\cdot s$.
Together, the new recurrence kernel $A\oplus A\ominus A^2\oplus B^2$ has a stride of $2s$,
 which doubles the stride of the original kernel $K$.

We can likewise analyze the number of taps in the new term.
If the original kernel $K$ has taps at indices $s,\ldots {k\cdot s}$, then the new recurrence kernel $A\oplus A\ominus A^2\oplus B^2$ also has $k$ taps in general, at indices $2s,\ldots,{2\cdot k\cdot s}$, and the new convolution kernel $I\ominus A\oplus B$ has $k+1$ taps.

If we apply dilation $n$ times, we obtain a new recurrence of the form $F = \rec(K_{n+1}, K_n\otimes \ldots \otimes K_1\otimes G)$
where $K_{n+1}$ has $k$ taps that are strided by $2^n$ and $K_i$ has $k+1$ taps for $i\neq n+1$.
This allows us to compute $2^n$ elements of the recurrence in parallel.

A \emph{delay} can further aid the dilation scheme. Consider the change of basis
\(\rec(A, G)= \rec(A\ominus B\otimes(I\ominus A),\ (I\oplus B)\otimes G)\)
again.
Let $B$ be the singleton kernel containing the first tap of $A$.
Then $A\ominus B\otimes (I\ominus A)$ effectively removes the first tap of $A$ while respecting the stride of $A$.
In other words, the new recurrence kernel keeps the same number of taps and the same stride, but widens the vectorizable work from $s$ to $2\cdot s$, as dilation does.
Unlike dilation, however, the new convolution kernel $I\oplus B$ has only two taps.
\begin{example}
    Consider the recurrence $F = \rec(A, G)$, where $A = [0, 0, a, 0, b, 0, c]$.
    Let $B=[0,0,a]$. Applying the change of basis, the rewritten recurrence kernel is
    $A\ominus B\otimes (I\ominus A)=[0,0,0,0,b+a^2,0,c+ab,0,ac]$,
    and the incurred convolution kernel is $I\oplus B=[1,0,a]$.
    This lets us compute four values of $F$ at a time, at the cost of an additional two-tap convolution.
    Dilating this kernel instead would yield a convolution with $I\ominus A\oplus B$, which has three taps.
\end{example}
Unlike dilation, applying delaying again
 does not double the vectorizable work,
 so it is best used as a replacement for the last step of dilation, saving taps in the convolution.
In practice, the work saved by delaying may be offset by the less regular vector-shuffling patterns, so our compiler tries both dilation and delaying as the last step and picks the faster implementation based on profiling.

\paragraph*{Idempotent recurrences}

To dilate time-invariant idempotent recurrences,
 we use \autoref{lem:idem-dilation}, which states that
    \(\rec(A\oplus B, G) = \rec(A\oplus B^2,\ (I\oplus B)\otimes G)\)
    for time-invariant kernels $A$ and $B$.
Generalizing this identity, we have
\(\rec\left(A\oplus \bigoplus_{i=1}^{n}B_i, G\right) = \rec\left(A\oplus\bigoplus_{i=1}^{n} B_i^2,\ \bigotimes_{i=1}^{n} (I\oplus B_i)\otimes G\right).\)
This identity lets us extract a set of singleton kernels $B_i$ from a kernel
 and dilate them independently.

Unlike the dilation for ring recurrences,
 which works only in the 1D case,
 the dilation for idempotent recurrences extends naturally to 2D.
For example,
 \autoref{exm:dist-trans} showed how dilation can expose opportunities for multi-core parallelism across rows.
Our compiler currently focuses only on vectorization and
 does not exploit such opportunities.

\subsection{General Recurrences}
\label{sec:general-recurrences}

We now describe techniques for vectorizing general recurrences.
Unlike the previous section, we assume neither time invariance nor any additional property of the semiring.
The two techniques below vectorize time-varying recurrences as well as time-invariant recurrences over semirings that lack additive inverses or are not idempotent.
Neither technique dominates the other,
 and the best choice depends on the target architecture.
For example, in our experience, Apple Silicon tends to prefer fast associative scan 
 while AMD Zen family tends to prefer path decomposition.

\subsubsection{Fast Associative Scan}
\label{sec:assoc-scan}

\citet{kogge-stone} first observed that an $m$-th order linear, time-invariant recurrence can be computed in parallel
 in logarithmic steps.
The key idea is a general transformation that parallelizes prefix scans of arbitrary associative operators \citep{BlellochTR90}; we show that it can also be used to vectorize recurrence relations.
To illustrate, let $F=\rec(K, G)$ where  $K$ is a $k$-tap kernel $[\algzero, S_1,\ldots, S_k]$.
Let us define $\;@\;$ as the matrix-vector and matrix-matrix multiply 
 using $\odot$, i.e., $(\mathcal{M}\;@\; \mathcal{N})_{i, j}=\bigoplus_{k}\mathcal{M}_{i, k}\odot \mathcal{N}_{k, j}$ and
$(\mathcal{M}\;@\; \mathcal{V})_{i}=\sum_{k}\mathcal{M}_{i, k}\odot \mathcal{V}_{k}$.
One step of the recurrence computation can be modeled as multiplication by a $(k+1)\times (k+1)$ matrix:
{\small
\begin{align*}
\begin{bmatrix}
    F\\
    \zk{-1}\otimes F\\
    \vdots\\
    \\
    \zk{-k+1} \otimes F\\
    \algone
\end{bmatrix}
=
\begin{bmatrix}
    S_1,&S_2,&\ldots, &S_k, &G\\
    \algone,& \algzero,&\ldots,&\algzero,&\algzero\\
    \algzero,& \algone,&\ldots,&\algzero,&\algzero\\
    \vdots&\vdots&\ddots &\vdots&\vdots\\
    \algzero,&\ldots, &\algone,&\algzero,&\algzero\\
    \algzero,&\ldots,&\algzero,&\algzero,&\algone\\
\end{bmatrix}
\;@\;
\begin{bmatrix}
    \zk{-1}\otimes F\\
    \zk{-2}\otimes F\\
    \vdots\\
    \\
    \zk{-k} \otimes F\\
    \algone
\end{bmatrix}
\end{align*}
}

Denote this $(k+1)\times (k+1)$ matrix by $\mathcal{M}$. Applying the transformation $n$ times, we obtain
\begin{align*}
\begin{bmatrix}
    F,
    \ldots,
    \zk{-k+1} \otimes F,
    \algone
\end{bmatrix}^{\top}
= \underbrace{\mathcal{M} \;@\; \Big(\ldots \;@\; \big(\mathcal{M}}_{n\text{ times}} \;@\;
\begin{bmatrix}
    \zk{-n}\otimes F,
    \ldots,
    \zk{-n-k+1} \otimes F,
    \algone
\end{bmatrix}^\top \big)\ldots\Big).
\end{align*}
The $@$ operator is associative, which follows because
 $\oplus$ distributes over $\odot$ and $\odot$ is associative.
Hence, the identity above can be written as
\(
[F,\ldots,\zk{-k+1} \otimes F, \algone]^\top = \mathcal{M}^n\;@\;[\zk{-n}\otimes F,\ldots,\zk{-n-k+1} \otimes F, \algone]^\top
\),
and the repeated $@$-product $\mathcal{M}^n$ can be computed in $\log n$ steps using the standard fast exponentiation trick (i.e., $\mathcal{M}^{2n}=\mathcal{M}^n @ \mathcal{M}^n$).

The power of $\mathcal{M}$ can be used for vectorization in different ways.
For example, on an architecture with $l$ vector lanes,
 a na\"ive approach computes $\mathcal{M}^{l}$ in $\log l$ steps
 and then computes $F$ as the dot product of the first row of $\mathcal{M}^l$ with $[\zk{-l}\otimes F,\ldots,\zk{-l-k+1} \otimes F, \algone]$.
Since the nearest self-dependency in the dot product is $\zk{-n}\otimes F$,
 vector instructions can compute $l$ values of $F$ at once.
In pseudocode, this is expressed as
\begin{lstlisting}[mathescape=true,basicstyle=\ttfamily\footnotesize]
f[x:x+l] = (${\bigoplus_{\tau}}_{i=1}^k$ $\left(\mathcal{M}^l\right)_{1, i}$[x:x+l] $\algtimes$ f[x-l-i+1:x-i+1]) $\oplus_\tau$ $\left(\mathcal{M}^{l}\right)_{1,k+1}$[x:x+l].
\end{lstlisting}
This form, however,
 uses only the first row of $\mathcal{M}^{l}$,
 even though each row of $\mathcal{M}^l$ can compute $F$ at a different index.
\looseness=-1 Ideally, with $l$ vector lanes, the power matrix should compute $k\times l$ outputs in parallel\footnote{We could instead compute the full $\mathcal{M}^{k\times l}$ directly and use its first row to produce $k\times l$ outputs, but this again uses only the first row and wastes work.}.


We propose a more efficient scheme that interleaves the output computation.
We compute only every $k$-th value of the power matrix $\mathcal{M}^{l\cdot k}$,
 which suffices to produce the full stream of $F$.
Specifically,
 we first de-interleave the stream of the initial tap matrix $\mathcal{M}$
 into $k$ streams, each produced at $1/k$ the rate of the original stream.
We then fold over the $k$ de-interleaved streams to compute $\mathcal{M}^k$,
 and use fast exponentiation to compute $\mathcal{M}^{2k}, \mathcal{M}^{4k},\ldots$.
All of $\mathcal{M}^k, \mathcal{M}^{2k}, \ldots$ are produced at rate $1/k$,
 so this does only $1/k$ the work of computing the full stream of $\mathcal{M}^{l\cdot k}$.
Finally, each matrix $\mathcal{M}^{l\cdot k}$
 computes $l$ values of $F$ in parallel, and we interleave these to produce the stream of
 output values $F$ at the original rate.
\autoref{fig:half-points} illustrates this computation scheme.
Our compiler uses the \texttt{WindowMap} operator to
 encapsulate this interleave/de-interleave scheme (\autoref{sec:compiler}).

\begin{figure}[ht]
    \centering
    {\scriptsize
 \begin{NiceTabular}{cccccccccc}[]
\textbf{Time} &$\ldots$ & $n-2k$ & $n-2k+1$ & $\ldots$ & $n-k$ & $n-k+1$ & $\ldots$ & $n-1$ & $n$ \\[2pt]
\textbf{Initial tap matrix} & $\ldots$ & $\mathcal{M}$ & $\mathcal{M}$ & $\ldots$ & $\mathcal{M}$ &$\mathcal{M}$ & $\ldots$ & $\mathcal{M}$ & $\mathcal{M}$ \\[2pt]
\textbf{Downsample by $k$} & $\ldots$ & $\mathcal{M}^k$ & & & $\mathcal{M}^k$ &  &  & & $\mathcal{M}^k$ \\[2pt]
\textbf{$2k$-parallel} & $\ldots$ & $\mathcal{M}^{2k}$ & & & $\mathcal{M}^{2k}$& &  &  & $\mathcal{M}^{2k}$ \\[2pt]
\textbf{$4k$-parallel} & $\ldots$ & $\mathcal{M}^{4k}$ & & & $\mathcal{M}^{4k}$ & &  &  & $\mathcal{M}^{4k}$ \\[2pt]
\textbf{$\ldots$} & $\ldots$ & $\ldots$ & & & $\ldots$ & &  &  & $\ldots$ \\[2pt]
\textbf{$l\cdot k$-parallel} & $\ldots$ & $\mathcal{M}^{l\cdot k}$ & & & $\mathcal{M}^{l\cdot k}$ & &  &  & $\mathcal{M}^{l\cdot k}$ \\[2pt]
\textbf{Output values $F$} & $\ldots$ & $F$ & $F$ & $\ldots$ & $F$ & $F$ & $\ldots$ & $F$ & $F$ \\
\CodeAfter
\tikz \draw[->] (2-2) -- (3-3);
\tikz \draw[->] (2-3) -- (3-3);
\tikz \draw[->] (2-4) -- (3-6);
\tikz \draw[->] (2-6) -- (3-6);
\tikz \draw[->] (2-7) -- (3-10);
\tikz \draw[->] (2-9) -- (3-10);
\tikz \draw[->] (2-10) -- (3-10);
\tikz \draw[->] (3-2) -- (4-3);
\tikz \draw[->] (3-3) -- (4-3);
\tikz \draw[->] (3-3) -- (4-6);
\tikz \draw[->] (3-6) -- (4-6);
\tikz \draw[->] (3-6) -- (4-10);
\tikz \draw[->] (3-10) -- (4-10);
\tikz \draw[->] (4-2) -- (5-3);
\tikz \draw[->] (4-3) -- (5-3);
\tikz \draw[->] (4-2) -- (5-6);
\tikz \draw[->] (4-6) -- (5-6);
\tikz \draw[->] (4-3) -- (5-10);
\tikz \draw[->] (4-10) -- (5-10);
\tikz \draw[->] (7-3) -- (8-2);
\tikz \draw[->] (7-3) -- (8-3);
\tikz \draw[->] (7-6) -- (8-4);
\tikz \draw[->] (7-6) -- (8-6);
\tikz \draw[->] (7-10) -- (8-7);
\tikz \draw[->] (7-10) -- (8-9);
\tikz \draw[->] (7-10) -- (8-10);
\end{NiceTabular}
    }
\caption{Vectorization using fast associative scan.}
\label{fig:half-points}
\end{figure}

\subsubsection{Path-based Decomposition}
\label{sec:path-decomp}

The second approach to vectorizing general recurrences rests on the following insight:
 a recurrence $F=\rec(K, G)$ computes the sums of all algebraic paths formed by the taps of $K$.
We can model this computation as a transition graph, in which each tap of the kernel $K$ is an edge, or a ``jump'', and the values of $G$ are the initial weights.
Specifically, the $k$-th tap of $K$ represents a jump of length $k$ and weight $\semof{K[i]}(n)$ at time $n$.

To formalize this notion, let $K_1,\ldots, K_k$ be the singleton kernels of $K$, each containing one tap of $K$.
Define $B_0=\vecone$ and $B_n=\bigoplus_{\substack{1\leq m\leq n\\a_1+\ldots+a_m=n\\\forall i.\ 1\leq a_i\leq k}} K_{a_1}\otimes \ldots\otimes K_{a_m}$
for $n\geq 1$.
Each $B_n$ is a singleton kernel with its tap at index $n$, and intuitively represents the sum of all algebraic paths of length $n$.
Define $G_p$ as the prefix scan of $B_0,\ldots, B_{p-1}$ convolved with the input signal $G$, i.e., $G_p=\bigoplus_{0\leq n < p} B_n \otimes G$.
Intuitively, $G_p$ aggregates all algebraic paths of length up to $p$, weighting each path by the sum of its weight and the initial weight from $G$.

\begin{example}
    Consider the kernel $K=[0, 2, 3]$ over the arithmetic ring.
    The singleton kernels are $K_1=[0, 2]$ and $K_2=[0, 0, 3]$.
    We have $B_0=[1]$, $B_1=K_1=[0,2]$, $B_2=K_1\otimes K_1 \oplus K_2=[0, 0, 3+2\times 2]=[0, 0, 7]$, 
    and 
    $B_3
    =[0, 0, 0, 2\times 3+3\times 2+2\times 2\times 2]
    =[0, 0, 0, 20]$.
    $B_3$ aggregates the combinations of $K$'s taps that sum to $3$:
    a two-jump followed by a one-jump ($K_1\otimes K_2$),
    a one-jump followed by a two-jump ($K_2\otimes K_1$),
    and three one-jumps ($K_1\otimes K_1\otimes K_1$).
    
    Let $G$ be an input signal $[1, -1, 1, -1]$.
    $G_4= (B_0 \oplus B_1 \oplus B_2 \oplus B_3)\otimes G=
    [1, 1, 6, 14]$.
    The $n$-th value of $G_4$ is the sum of paths ending at $n$.
    For instance, $\semof{G_4}(3)=14$, the sum of 0-paths ($1\times -1$),
    1-paths ($2\times 1$), 2-paths ($7\times -1$), and 3-paths ($20\times 1$).
\end{example}

The following lemma captures the intuition that the recurrence $F=\rec(K, G)$ computes the sum of algebraic paths of every length.

\begin{lemma} 
Let $K$ be a kernel and $G$ be a signal over any semiring.
Define $F=\rec(K, G)$. We have $F = \bigoplus_{n\geq 0} B_n \otimes G$.
\end{lemma}

The lemma follows by induction on the time index $\veci$,
 and the unbounded sum is well-defined by the same argument as for the well-definedness of  $\rec(K, G)$.

Under this path-based interpretation, the following lemma states
 that we can decompose the computation of $F$ into two parts:
 a local contribution from paths of length less than $p$,
 and a global contribution from paths that start outside the local $p$-sized window.
\looseness=-1 Only the global contribution has far self-references to $F$.

\begin{lemma}
    \label{lem:path-decomp}
For any $F=\rec(K, G)$, we have
{\small
\[
    F=
    G_p
    \oplus 
    \left(
    \bigoplus\nolimits_{1\leq q\leq k}
    \bigoplus\nolimits_{0\leq r<p,\ p\leq r+q}
     B_r\otimes K_q\otimes 
    F
    \right).
\]
}
\end{lemma}

In this lemma, $G_p$ gives the local contribution,
 and each global contribution has the form $B_r\otimes K_q\otimes F$: $B_r$ aggregates the paths
 that traverse the local window, $K_q$ is the jump that leaves the local window, and $F$ is the continuation of the path.
To enumerate the global contributions, we classify paths by the first jump out of the local window:
 we first enumerate $q$, the length of that jump, and then $r$, the position where it is made.
The position $r$ must lie within the local window, i.e., $0\leq r<p$, and the jump must leave it, i.e., $p\leq r+q$.

The following two lemmas show how to divide-and-conquer the computation of $B_n$ and $G_p$.
Both take a similar form, classifying paths by the first jump out of the local window.

\begin{lemma}
    \label{lem:path-decomp-B}
    For any $n, p$ where $p< n$,
\(
B_n=\bigoplus\nolimits_{1\leq q\leq k} \bigoplus\nolimits_{0\leq r<p,\ p\leq r+q\leq n}{B_r}\otimes K_q\otimes {B_{n-r-q}}.
\)
\end{lemma}

\begin{lemma}
    \label{lem:path-decomp-G}
    For any $p, l$ where $l<p$,
\(G_p=
   G_l
   \oplus
   \left(
   \bigoplus\nolimits_{1\leq q\leq k} 
   \bigoplus_{\substack{0\leq r< l\\ l\leq  r+q}}
   {B_r}\otimes K_q\otimes 
   G_{p-r-q}
   \right).\)
\end{lemma}

\begin{figure}[t]
    \centering\includegraphics[width=0.8\textwidth]{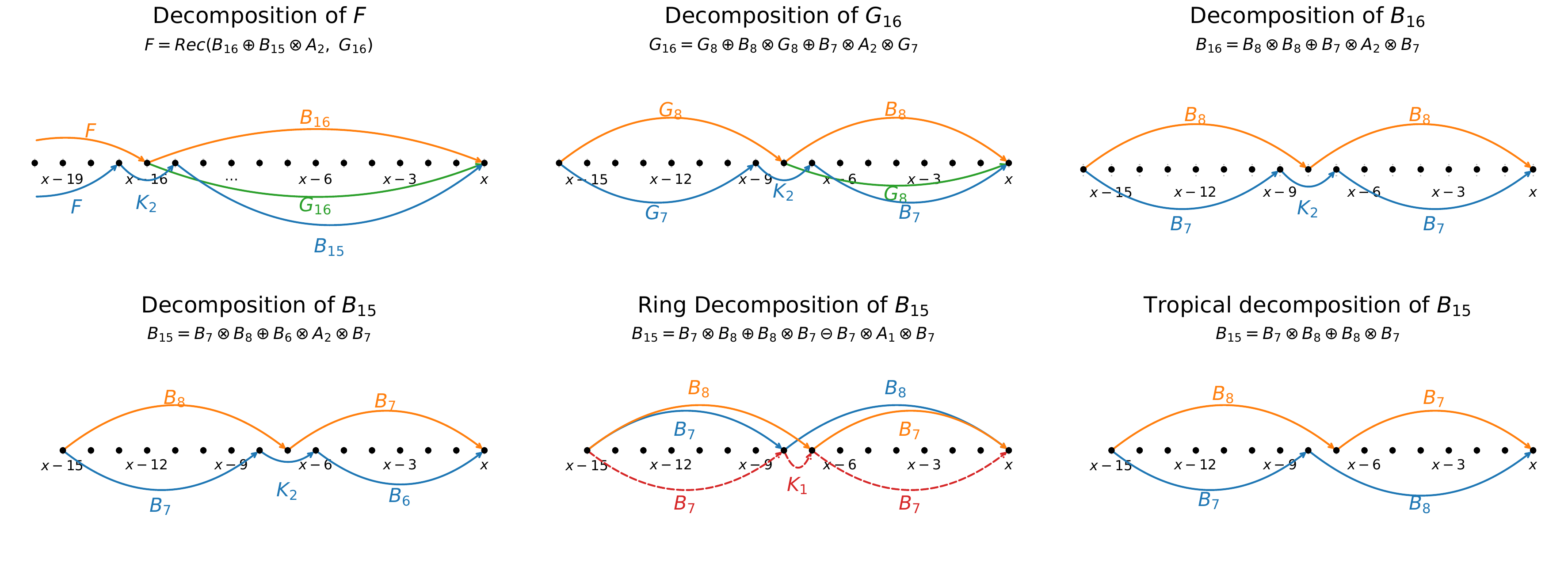}
    \caption{Visual illustration of path decompositions for $F,G_{16}, B_{16}, B_{15}$.
    Dashed line denotes a subtracted path.}
    \label{fig:path-decomp}
\end{figure}

Taken together, these lemmas let us compute the recurrence $F$ in terms of $B_n$ and $G_n$,
 and divide-and-conquer $B_n$ and $G_p$ into smaller pieces.
Below, we give examples illustrating how to use these lemmas to vectorize $F$.
\autoref{fig:path-decomp} visualizes the used path decompositions.

\begin{example}
    \label{ex:vanilla-path-decomp}
    Consider a second-order time-varying recurrence $F=\rec(A, G)$ where $A=[\algzero, S_1, S_2]$.
    Define the singleton kernels
     $A_1=[\algzero, S_1]$,
     and $A_2=[\algzero,\algzero,S_2]$.
    We can compute $F$ in terms of $G_{16},B_{16}, B_{15}$:
    {\small
    \begin{align*}
    F &=\rec(\ B_{15}\otimes A_1\;\oplus\; B_{14}\otimes A_2 \;\oplus\; B_{15}\otimes A_2,\  G_{16}\ )\\
        &=\rec(\ B_{16}\;\oplus\; B_{15}\otimes A_2,\  G_{16}\ )
    \end{align*}
    }
    This uses the identity $B_i = B_{i-1}\otimes A_1 \oplus B_{i-2}\otimes A_2$
    to simplify the computation.
    The corresponding visualization shows that all algebraic paths of $F$ can be classified as
     paths from the local window (i.e., $G_{16}$), paths hitting $x\compactsub16$ (i.e., $B_{16}\otimes F$), and paths skipping $x\compactsub16$ via a two-jump (i.e., $B_{15}\otimes K_2\otimes F$).
    Similarly, we can compute $G_{16}, B_{16}, B_{15}$ from $B_8, B_7, B_6, G_8, G_7$.
    {\small
    \begin{align*}
        G_{16} 
        &= G_8\;\oplus\; B_{8}\otimes G_8\;\oplus\; B_{7}\otimes A_2\otimes G_7\\
        B_{16} 
        &=\quad\quad \ \ \ B_{8}\otimes B_8\;\oplus\; B_{7}\otimes A_2\otimes B_7\\
        B_{15} 
        &= \quad\quad \ \ \  B_{7}\otimes B_8\;\oplus\; B_{6}\otimes A_2\otimes B_7
        \tag*{\qedhere}
    \end{align*}
    }
    Take $G_{16}$ for example. The visualization suggests the local paths that $G_{16}$ aggregates can be categorized into three classes: from a local window of $G_8$, from outside the window and hitting $x\compactsub8$, and from outside the window and skipping $x\compactsub8$ via a two-jump.
\end{example}

\begin{example}
    \label{ex:ring-path-decomp}
    If we specialize the recurrence to rings,
     we can use the fact that $B_{6}\otimes A_2\otimes B_7 = B_8\otimes B_7\;\ominus\; B_{7}\otimes A_1\otimes B_7$ to eliminate the dependence on $B_6$.
     {\small
     \begin{align*}
        B_{15}
        &= B_{7}\otimes B_8\;\oplus\; B_8\otimes B_7\;\ominus\; B_{7}\otimes A_1\otimes B_7
        \tag*{\qedhere}
    \end{align*}
    }
    This saves the work of computing $G_6$ and its dependencies.
    Intuitively, the decomposition says that to aggregate over all paths of length $15$, we can consider paths that hit $x\compactsub7$ and paths that hit $x\compactsub8$, and then subtract paths that hit both $x\compactsub7$ and $x\compactsub8$.
\end{example}

\begin{example}
    If we specialize the recurrence to idempotent semirings instead,
    we can again avoid $B_6$, this time by relying on idempotence and intentionally double-counting
    the path $B_7\otimes A_1\otimes B_7$.
    {\small
    \begin{align*}
        %
        B_{15}
        &= B_{7}\otimes B_8\;\oplus\; B_8\otimes B_7
        \tag*{\qedhere}
    \end{align*}
    }
\end{example}

There are several ways to divide and conquer a recurrence, and algebraic identities can rearrange it into forms that save even more work.
The general forms in \autoref{lem:path-decomp-B} and \autoref{lem:path-decomp-G} are often not the most efficient, as they yield quadratically many terms and can introduce too many dependencies.
Our compiler hard-codes the optimal decomposition for orders of 2 or fewer over rings and idempotent semirings, which covers the most common recurrences.
Scaling to higher order recurrences is left as future work.

\subsection{Extensions}
\label{sec:extension}

We now extend the techniques of \autoref{sec:vectorization-tricks}
 to two-dimensional and mutually recursive recurrences.

\paragraph{Two-dimensional recurrences}

\emph{Marginalization} lets us treat a higher-dimensional recurrence
 as a one-dimensional one.
For example, consider the 2D recurrence $f(x, y) = f(x\compactsub 1, y) + f(x, y\compactsub 1) + g(x, y)$.
Although $f(x, y)$ depends on itself in the preceding row and column,
 we can marginalize $f(x, y)$ (resp.~$g(x, y)$) over $x$
 into a family of 1D signals $f_x(y)$ (resp.~$g_x(y)$) indexed by $x$.
This gives $f_x(y) = f_x(y\compactsub 1) + f_{x-1}(y) + g_x(y)$,
 which is a prefix sum of $f_{x-1}(y)+g_x(y)$ and we know how to vectorize.
This also matches how 2D recurrences are computed in practice:
 row by row, where all earlier rows have been computed when computing the current one.

In \autoref{sec:compiler} we discuss how to generate code for 2D recurrences
 using a repeater operator $\texttt{Repeater }\mu x. e$,
 which lets $e$ refer to already-computed rows via $x$.

\paragraph{Mutual recurrences}

Mutual recurrences are a group of recurrences that depend on each other.
For example, double exponential smoothing is a method that computes the interdependent smoothed value and trend value of a time series,
 and it can be expressed as a mutual recurrence:
{\small
\begin{align*}
l(x) &= (1-\alpha) \cdot l(x\compactsub 1) \ +\ \ (1-\alpha) \ \cdot b(x\compactsub 1) \ +\  \ \alpha\ \:\cdot g(x)\\
b(x) &= \ \ -\alpha\beta\ \ \:\cdot l(x\compactsub 1) \ +\  (1-\alpha\beta) \cdot b(x\compactsub 1) \ +\  \alpha\beta \cdot g(x)
\end{align*}
}
%
The LeetCode problem from the introduction is also mutually recursive.

Mutual recurrences fit naturally into our framework: we lift signals to vectors of signals,
 and kernels to matrices of kernels.
For instance, the double exponential smoothing recurrence can be expressed as a recurrence vector:
{\small
\[
\begin{bmatrix}
    L\\ B
\end{bmatrix} = 
\begin{bmatrix}
    \begin{bmatrix}\scriptstyle 0 & \scriptstyle 1-\alpha \end{bmatrix} &
    \begin{bmatrix}\scriptstyle 0 & \scriptstyle 1-\alpha \end{bmatrix} \\[0.6em]
    \begin{bmatrix}\scriptstyle 0 & \scriptstyle -\alpha\beta \end{bmatrix} &
    \begin{bmatrix}\scriptstyle 0 & \scriptstyle 1-\alpha\beta \end{bmatrix}
\end{bmatrix}
\otimes 
\begin{bmatrix}
    L\\ B
\end{bmatrix}
\oplus
\begin{bmatrix}
    \alpha\cdot G\\
    \alpha\beta\cdot G
\end{bmatrix}
\]
}

The same algebraic transformations and vectorizing techniques that work for standard recurrences
 also apply to recurrence vectors.
Matrix multiplication, however, is not commutative in general,
 so we no longer have $A\otimes B=B\otimes A$ for kernel matrices $A$ and $B$,
 even when the kernels within $A$ and $B$ are time-invariant.
Therefore, we can apply only those vectorizing transformations
 that do not rely on commutativity, which excludes dilation.


\section{The \vecrec{} Compiler}
\label{sec:compiler}






In this section, we introduce \vecrec{}, our compiler instantiation of the recurrence algebra.
The recurrence algebra itself leaves many details of physical implementation unspecified.
We first describe \vecrec{}'s domain-specific language (DSL), which exposes
 details relevant to physical execution, and then the codegen process
 that lowers the DSL into efficient vectorized \CC{} implementations.

\begin{figure}[t]
  \centering
  \begin{lstlisting}[basicstyle=\ttfamily\scriptsize]
type Kernel1D, Signal1D, Kernel2D, Signal2D

TimeInv : [Number] -> Kernel1D;    TimeInv2D : [[Number]] -> Kernel2D
TimeVar : [Signal1D] -> Kernel1D;  TimeVar2D : [[Signal1D]] -> Kernel2D
Var : String -> Signal1D;          Var2D : String -> Signal2D

Convolve   : (Kernel1D, Signal1D) -> Signal1D
Recurse    : (Kernel1D, Signal1D) -> Signal1D
Convolve2D : (Kernel2D, Signal2D) -> Signal1D
Recurse2D  : (Kernel2D, Signal1D) -> Signal2D

WindowMap : (n: Number, [Signal1D], [[Signal1D; n]] -> Signal1D) -> Signal1D

Repeater : (Signal2D -> Signal1D) -> Signal2D
Ith      : (Signal2D, Int) -> Signal1D

ConvertLanes : forall T. (T, Int) -> T
\end{lstlisting}
\caption{The domain-specific language used in \vecrec{}. The full version is available in the appendix of the full version.}
\label{fig:reclang-syntax}
\end{figure}

\vecrec{} uses a streaming computation model and exposes a \texttt{run} method that, when called,
 returns the current output vector(s) and advances the internal state.
This interface assumes that neither the input nor the output is stored in memory;
 instead, it stores internally only the vectors needed to compute the output vectors.

\subsection{\vecrec{}'s Domain-Specific Language}

\autoref{fig:reclang-syntax} shows a fragment of the \vecrec{} syntax;
the full syntax is available in the appendix of the full version.
Here we omit some pointwise operators and the syntax for mutual recurrences.
\rev{Mutual recurrences are supported with kernel matrix types and signal vector types,
 and operators like \texttt{Convolve} and \texttt{Recurse} are overloaded to accept kernel matrices and signal vectors.}

Unlike in the recurrence algebra, signal and kernel types in \vecrec{} are refined by dimension.
\vecrec{} considers only one- and two-dimensional kernels and signals,
 which suffice for our applications.
\texttt{Kernel1D} and \texttt{Kernel2D} represent a list of taps and a matrix of signal taps, respectively,
 and each tap is a 1D signal.
The signal types are more interesting.
The one-dimensional signal type \texttt{Signal1D} streams one value at a time,
 while the two-dimensional signal type \texttt{Signal2D} streams a tuple of not only the current value
 but also the values from the previous rows in the same column.
To motivate this distinction, consider the convolution $A\otimes G$ for a 2D kernel $A$.
To compute its value at a point $(x, y)$, we must fetch the values of $G$ at not only the current row $x$
 but also the previous rows $x\compactsub 1, x\compactsub 2, \ldots$, based on the window defined by $A$.
In the streaming model, we must buffer the last $n$ rows of $G$ when
 $G$ is not already materialized (e.g., a 2D input signal).

\texttt{TimeInv} (resp.\ \texttt{TimeVar}) constructs 1D time-invariant (resp.\ time-varying) kernels
 and takes numbers (resp.\ 1D signals) as inputs.
\texttt{TimeInv2D} and \texttt{TimeVar2D} are the analogues for 2D kernels.
\texttt{Var} and \texttt{Var2D} construct 1D and 2D input signals.

\texttt{Convolve} and \texttt{Recurse} express 1D convolutions and recurrences,
 and they take and return 1D signals.
The 2D convolution and recurrence operators have interesting type signatures.
\texttt{Convolve2D} takes a 2D signal but returns a 1D signal.
This is because a physical implementation of convolution reads
 a 2D window of rows from the input but does not buffer by itself.
\texttt{Recurse2D}, in contrast, takes a 1D signal but returns a 2D signal.
This is because a 2D recurrence convolves with its own outputs from previous rows
 and must cache these rows.
Meanwhile, it does not require the input signal to be buffered (hence a 1D input signal).


A \texttt{Repeater} turns a procedure that produces 1D signals into a 2D signal.
It computes the given 1D signal one row at a time,
 buffers the rows already produced, and feeds them as non-recursive inputs to the current row.
$\texttt{Repeater}(\lambda s.\,e)$ binds $s$ to the 2D signal that represents the previously computed rows,
 which $e$ uses to compute the current row.
The \texttt{Repeater} expression returns a 2D signal that streams both the current row and the buffered rows.
\texttt{Ith} destructs a 2D signal and extracts a 1D signal at a given row offset.
Together they enable marginalization and let us express arbitrary 2D recurrences as 1D recurrences.
In fact, \vecrec{} desugars a \texttt{Recurse2D} into a \texttt{Repeater} expression:
\(
  \texttt{Recurse2D}(K, S) \equiv 
  \texttt{Repeater}(\lambda S'.\, \texttt{Recurse}(K_0, \texttt{Add}(\texttt{Convolve}(K_1, \texttt{Ith}(S', 0)), \ldots))),
\)
where $K_i$ is a 1D kernel representing the $i$-th row of the 2D kernel $K$.

\texttt{WindowMap} de-interleaves lists of signals and is used by the fast associative scan.
Given interleave factor $n$, it splits each input into $n$ substreams, where substream $j$ takes every $n$-th sample
 from offset $j$; it then runs its body on the substreams and interleaves the results back to full rate.

Each operator in \vecrec{} is implicitly annotated with
 the number of lanes it operates over at a time.
By default, an operator is annotated with the maximum number of vector lanes available to the hardware.
The number of vector lanes in a recurrence, however, is bounded by the index of the first non-zero tap of the kernel.
This bound can cause the vector lanes of a node to differ from those of its child or parent.
We use \texttt{ConvertLanes(e, l)} to explicitly convert
 the vector lanes of $e$ to $l$ lanes.
The number of lanes of $e$ must be either a divisor or a multiple of $l$.
In the physical implementation, if $e$ has fewer lanes than the specified $l$,
we repeat the physical execution of $e$ for $l/l_e$ times, where $l_e$ is the number of lanes of $e$.
Otherwise, we compute $e$ on every $l_e/l$-th invocation of the operator
 and split the output vector among $l_e/l$ calls.

\subsection{Compiler Implementation}

We implement the recurrence identities of \autoref{sec:algebra} and
 the vectorizing transformations of \autoref{sec:vectorization-tricks} as compiler passes.
These passes enumerate combinations of transformations and vectorization factors
 to produce a set of distinct vectorized forms. We call each form a \emph{schedule}.
Among the set of schedules, the compiler uses runtime profiling to select the best one.

Given a schedule, the code generator must first infer (1) the number of buffered rows for 2D signals and (2) the number of lanes for each operator.
We obtain the number of buffered rows needed for a 2D signal by collecting the \texttt{Ith} references to it and taking the maximum.
For vector lane annotations,
 we use the maximum available lanes for non-recurrence nodes.
For recurrence operators,
 the vector lanes are given by the position of the first non-zero tap.
If the vector lane annotation of a node does not match that of its child,
 we fork the expression into two schedules:
 either we wrap the child with \texttt{ConvertLanes}, so that the child's annotation remains unchanged,
 or we recursively downgrade the vector lane annotations of the children to match the parent.
\begin{example} 
Consider the recurrence \texttt{Recurse(K1, Convolve(K2, Convolve(K3, g)))},
 where the first non-zero tap of \texttt{K1} is at offset 8
 and the hardware supports up to 16 vector lanes.
Initially, the \texttt{Recurse} node is annotated with 8 lanes and the two \texttt{Convolves} with 16 lanes,
 so the lanes of \texttt{Recurse} and \texttt{Convolve} differ.
The compiler generates two versions of the annotations. In the first, it inserts a \texttt{ConvertLanes} node between \texttt{Recurse} and \texttt{Convolve} that converts from 16 lanes to 8 lanes.
In the second, it recursively pushes \texttt{Recurse}'s annotation down to its descendants.
Both versions are considered for runtime profiling.
However, the compiler never inserts a \texttt{ConvertLanes} node between two non-recursive nodes (e.g., between two \texttt{Convolve} nodes).
\end{example}

Once a schedule is fully annotated with buffered rows and vector lanes,
 we lower it into a \CC{} struct.
The code generator keeps a \CC{} template library of operator structs such as \texttt{Recurse} and \texttt{Convolve}.
These structs operate over a register state struct, which wraps an array of vector types\footnote{We use \texttt{clang}'s vector extensions~\citep{clang_language_extensions}.}.
Operators read input vectors and write output vectors at given offsets of this array,
 and maintain internal state. 
For example, a \lstinline{Convolve} keeps a window of the last $n$ input vectors that is enough for the next convolution, and a \lstinline{Recurse} similarly keeps previously computed output vectors.
The code generator traverses the tree bottom up and emits an operator for each distinct node.
The resulting \CC{} struct is initialized and called from user code.

\subsection{Determining Optimal Unrolling Factors}
\label{sec:unrolling}

For each generated kernel program, we measure its throughput on random input signals.
For the top 10\% of kernels, we perform an additional sweep to find the optimal unrolling factors.
A \vecrec{} program can be unrolled horizontally and, for 2D recurrences, vertically.
An unrolled program produces a tile of output vectors at a time.
For 2D recurrences, vertical unrolling saves memory roundtrips that horizontal unrolling cannot,
 because the vertical dependencies already reside in registers and need not be loaded from memory\footnote{
  Horizontal unrolling saves no memory roundtrips
  because 1D operators always keep their internal state in vector registers.
 }.
Not all 2D recurrences benefit from vertical unrolling on all architectures, however,
 because vectorized kernels are often bound by register pressure
 (from the introduced convolutions) rather than memory bandwidth.
It helps most on ARM, which offers more instruction-level parallelism.


\section{Evaluation}
\label{sec:evaluation}

We evaluate \vecrec{} in two parts. 
First, we benchmark \vecrec{}-generated kernels against real-world recurrence applications and LeetCode DP problems. 
\rev{
    To the best of our knowledge, most real-world recurrence implementations do not vectorize\footnote{This inlcudes scipy.sosfilt, DSP.jl, ffmpeg, torchaudio, Faust.}, except for Intel IPP and Apple Accelerate,
    which have vectorized implementations for biquad filters.
    Therefore, we compare against Intel IPP and Apple Accelerate on biquad filters.
    For the rest of real-world applications and all LeetCode problems, we compare against performant, unvectorized baselines.
    In both cases, we demonstrate significant speedups.
}
Second, we conduct a case study of bioinformatic recurrences, which many studies have sought to speed up with vectorization~\cite{parasail,striped,vec-sequence-comparison,Rognes2011FasterSD}.
To that end, we compare the kernels with vectorized implementations from state-of-the-art libraries, and integrate them into existing systems, showing up to 28\% speedup on end-to-end pipelines.
Together, our evaluation shows \vecrec{} works on a broad range of recurrences,
beats hand-tuned algorithms for specific ones, and provides state-of-the-art performance on diverse hardware.

We evaluate \vecrec{} on three platforms that span different processor generations, microarchitectures, and instruction sets:
 Intel Skylake (Core i9-9960X, 2018),  AMD Zen 5 (Ryzen Threadripper 9970X, 2025), and Apple Silicon (M2, 2022).
The first two architectures support the AVX-512 instruction set with 512-bit operations. The M2's NEON instruction set supports only 128-bit operations, but it offers better instruction-level parallelism and can issue more instructions simultaneously.

\subsection{Evaluation on Recurrences in the Wild}

\rev{
Most high-performance libraries for signal processing just run recurrences as scalar code,
 so we compare \vecrec{}-generated kernels with Intel IPP and Apple Accelerate, two libraries with vectorized implementations for biquad filters.
Both libraries's vectorization schemes incur the additional work that grows linearly with the exposed parallelism,
 while in ours it grows only logarithmically.
Both libraries vectorize biquad filters with a fixed width of 4.
The result is shown in \autoref{tab:biquad-throughput}. \vecrec{} is faster than both vendor libraries by up to $5.38\times$.
}

\begin{table}[h]
\caption{Speedup over baselines from vendor libraries on biquad filters. We use IPP's \texttt{ippsIIR\_32f} and Apple Accelerate's \texttt{vDSP\_deq22} for the comparison.}
\label{tab:biquad-throughput}
\resizebox{0.8\textwidth}{!}{%
\centering
\footnotesize\setlength{\tabcolsep}{4pt}{%
\begin{tabular}{l|rrr|rrr|rrr}
\toprule
Benchmark & \multicolumn{3}{c}{Skylake} & \multicolumn{3}{c}{Zen 5} & \multicolumn{3}{c}{Apple M2} \\
\cmidrule(lr){2-4} \cmidrule(lr){5-7} \cmidrule(lr){8-10}
 & Intel IPP & Vec & Speedup & Intel IPP & Vec & Speedup & Apple Accelerate & Vec & Speedup \\
 & \multicolumn{2}{c}{(Melem/s)} & (\(\times\)) & \multicolumn{2}{c}{(Melem/s)} & (\(\times\)) & \multicolumn{2}{c}{(Melem/s)} & (\(\times\)) \\
\midrule
Biquad Filter & 770.4 & 2111.8 & 2.74 & 1168.5 & 6282.3 & 5.38 & 1827.8 & 3063.2 & 1.68 \\
\bottomrule
\end{tabular}
}
}
\end{table}

\rev{Next, we evaluate on a large class of recurrences that arise naturally in signal processing and numerical computing,
 many of which are critical in their respective domains.
Here we highlight two rather unexpected applications of recurrences:

}

\begin{itemize}[leftmargin=*,nosep]
    \item Continued fraction: A continued fraction has the form \(b_0+a_1/(b_1+a_2/(\ldots))\) and can be computed using the Wallis-Euler recurrence, defined by
    {\small
    \begin{align*}
        h(x) &= a(x)\cdot h(x\compactsub 1) + b(x)\cdot h(x\compactsub 2)\\
        f(x) &= h(x)/h(x\compactsub 1).
    \end{align*}
    }
    As $x$ goes to infinity, $f$ converges to the continued fraction. An example of a continued fraction is to use the ratio of two consecutive Fibonacci numbers to approximate the golden ratio: $1/(1+1/(1+\ldots))$.
   In general the coefficients are time-varying, and
    \vecrec{} can vectorize such recurrences.
    \item Cumulative variance: Given a list of values $a$, Welford's algorithm~\citep{Welford1962NoteOA} 
    computes the variances of all prefixes of $a$ in linear time.
    It can be defined as the following cascade of recurrences:
    {\small \begin{align*}
        \textit{Mean}(x) \ &=\  (1-1/x)\cdot \textit{Mean}(x\compactsub 1) \ +\  1/x\cdot a(x)\\
        M_2(x)\ &=\ M_2(x\compactsub 1)\ +\ (a(x) - \textit{Mean}(x))\cdot (a(x)-\textit{Mean}(x\compactsub 1))\\
        \textit{Var}(x) \ &=\  1/x \cdot M_2(x).
    \end{align*} }
    \vecrec{} can represent this recurrence and generate an efficient vectorized kernel, as $\textit{Mean}$ is a first-order time-varying recurrence and $M_2$ is a prefix scan.
\end{itemize}

\begin{table}[t]
\caption{Benchmarked throughputs for recurrence applications on the serial, baseline, and vectorized implementations across architectures. Speedup is Vec over the faster of Serial and Base.}
\label{tab:recurrence-throughput}
\resizebox{\textwidth}{!}{%
\centering
\small
\begin{tabular}{l|rrrr|rrrr|rrrr}
\toprule
Benchmark & \multicolumn{4}{c}{Skylake} & \multicolumn{4}{c}{Zen 5} & \multicolumn{4}{c}{Apple M2} \\
\cmidrule(lr){2-5} \cmidrule(lr){6-9} \cmidrule(lr){10-13}
 & Serial & Base & Vec & Speedup & Serial & Base & Vec & Speedup & Serial & Base & Vec & Speedup \\
 & \multicolumn{3}{c}{(Melem/s)} & (\(\times\)) & \multicolumn{3}{c}{(Melem/s)} & (\(\times\)) & \multicolumn{3}{c}{(Melem/s)} & (\(\times\)) \\
\midrule
First-Order High Pass Filter & 331.9 & 417.5 & 2320.9 & 5.56 & 676.9 & 676.8 & 9510.6 & 14.05 & 349.9 & 348.4 & 4678.5 & 13.37 \\
Second-Order High Pass Filter & 248.9 & 396.1 & 2020.1 & 5.10 & 403.4 & 672.3 & 5939.3 & 8.83 & 205.4 & 316.8 & 2564.3 & 8.09 \\
Second-Order Band-Pass Filter & 497.0 & 440.4 & 2186.6 & 4.40 & 773.4 & 1098.9 & 7366.4 & 6.70 & 497.0 & 717.3 & 3016.6 & 4.21 \\
Box Filter & 248.2 & 239.8 & 1967.9 & 7.93 & 1032.4 & 539.6 & 6278.9 & 6.08 & 431.5 & 416.9 & 2960.7 & 6.86 \\
Repeated Box Filter & 89.9 & 92.6 & 852.2 & 9.20 & 246.7 & 235.7 & 2037.1 & 8.26 & 119.0 & 175.6 & 896.3 & 5.10 \\
Recursive Gaussian Filter & 248.9 & 249.0 & 1884.9 & 7.57 & 361.0 & 361.1 & 5721.9 & 15.85 & 218.9 & 194.4 & 2507.1 & 11.45 \\
Non-uniform Recursive Filter & 483.0 & 482.6 & 865.3 & 1.79 & 676.3 & 720.9 & 2761.0 & 3.83 & 317.0 & 350.2 & 1656.4 & 4.73 \\
Distance Transform & 281.9 & 268.8 & 1680.5 & 5.96 & 548.3 & 409.7 & 5038.8 & 9.19 & 272.0 & 427.9 & 1985.6 & 4.64 \\
Exponential Smoothing & 497.3 & 736.3 & 2431.3 & 3.30 & 773.6 & 773.7 & 10217.1 & 13.21 & 437.9 & 697.3 & 4873.3 & 6.99 \\
Double Exponential Smoothing & 199.1 & 411.0 & 1691.5 & 4.12 & 361.1 & 545.8 & 4368.8 & 8.01 & 205.5 & 346.5 & 1880.2 & 5.43 \\
Welford Variance & 438.8 & 173.0 & 1404.7 & 3.20 & 676.0 & 354.3 & 3708.6 & 5.49 & 248.7 & 250.1 & 1778.3 & 7.11 \\
Continued Fraction & 412.3 & 386.1 & 429.5 & 1.04 & 659.1 & 686.7 & 1505.6 & 2.19 & 276.6 & 456.0 & 1087.9 & 2.39 \\
\bottomrule
\end{tabular}
}
\end{table}

\begin{table}[t]
\caption{Benchmarked throughputs on LeetCode problems.}
\label{tab:leetcode-throughput}
\resizebox{\textwidth}{!}{%
\centering
\small
\begin{tabular}{l|rrrr|rrrr|rrrr}
\toprule
Benchmark & \multicolumn{4}{c}{Skylake} & \multicolumn{4}{c}{Zen 5} & \multicolumn{4}{c}{Apple M2} \\
\cmidrule(lr){2-5} \cmidrule(lr){6-9} \cmidrule(lr){10-13}
 & Serial & Base & Vec & Speedup & Serial & Base & Vec & Speedup & Serial & Base & Vec & Speedup \\
 & \multicolumn{3}{c}{(Melem/s)} & (\(\times\)) & \multicolumn{3}{c}{(Melem/s)} & (\(\times\)) & \multicolumn{3}{c}{(Melem/s)} & (\(\times\)) \\
\midrule
10.~Regular Expression Matching & 735.2 & 183.9 & 1772.8 & 2.41 & 1499.0 & 488.9 & 4405.0 & 2.94 & 1185.0 & 331.9 & 1862.5 & 1.57 \\
91.~Decode Ways & 719.1 & 394.4 & 542.7 & 0.75 & 1086.1 & 1080.0 & 3255.6 & 3.00 & 504.0 & 738.6 & 1870.9 & 2.53 \\
121.~Buy/Sell Stock & 1450.8 & 544.5 & 2187.9 & 1.51 & 1792.8 & 983.8 & 6907.1 & 3.85 & 1606.6 & 1455.7 & 2667.2 & 1.66 \\
188.~Buy/Sell Stock IV & 437.3 & 516.9 & 2158.2 & 4.18 & 844.9 & 1257.8 & 4336.8 & 3.45 & 404.1 & 717.5 & 2432.2 & 3.39 \\
198.~House Robber & 637.7 & 591.4 & 1169.7 & 1.83 & 978.8 & 898.6 & 3676.9 & 3.76 & 447.5 & 1409.9 & 1702.0 & 1.21 \\
221.~Maximal Square & 356.0 & 387.2 & 2182.3 & 5.64 & 772.8 & 415.1 & 6688.7 & 8.66 & 701.4 & 211.2 & 2713.0 & 3.87 \\
304.~Range Sum Query 2D & 1557.0 & 646.1 & 2327.9 & 1.50 & 2679.3 & 883.0 & 6406.4 & 2.39 & 1694.1 & 2080.4 & 3943.9 & 1.90 \\
309.~Buy/Sell Stock with Cooldown & 704.1 & 529.8 & 686.8 & 0.98 & 1019.5 & 1149.4 & 1709.9 & 1.49 & 597.0 & 740.0 & 944.6 & 1.28 \\
746.~Min Cost Climbing Stairs & 599.0 & 200.6 & 1206.4 & 2.01 & 983.2 & 308.8 & 2898.7 & 2.95 & 439.2 & 704.1 & 980.6 & 1.39 \\
1143.~Longest Common Subsequence & 282.8 & 714.6 & 2330.8 & 3.26 & 378.2 & 1414.3 & 6682.2 & 4.72 & 1092.4 & 803.2 & 3291.3 & 3.01 \\
1458.~Max Dot Product of Two Subsequences & 540.1 & 602.6 & 2312.6 & 3.84 & 901.6 & 1090.2 & 9837.0 & 9.02 & 1066.3 & 761.6 & 3065.4 & 2.87 \\
3259.~Max Energy Boost From Two Drinks & 213.7 & 282.3 & 549.0 & 1.94 & 653.4 & 513.5 & 1382.5 & 2.12 & 297.6 & 852.0 & 881.0 & 1.03 \\
\bottomrule
\end{tabular}
}
\end{table}

For these, our baselines are `Serial', a manual \CC{} implementation that computes the recurrences in serial loops, and `Base', a kernel that \vecrec{} produces but with vectorization disabled.
We set the problem size to $2^{22}$ and measure the throughput.
\autoref{tab:recurrence-throughput} shows the runtime on recurrences that arise naturally in signal processing and numerical computing.
Base sometimes achieves higher throughput than Serial because \vecrec{} unrolls the recurrence computation using a runtime profile.
Serial can be faster if it hits a better compilation path in \texttt{clang}.
The speedup is reported as \vecrec{} over the higher throughput of Serial and Base.

Overall, \vecrec{} achieves a speedup of 1.04$\times$ to 15.85$\times$.
The speedup tends to be bigger for time-invariant recurrences where dilation applies.
\rev{Additionally, many benchmarks, including continued fraction and Welford variance mentioned above,
 consist of both recurrence operators and expensive pointwise operators like float division.
Recurrence's self-dependencies prevent the rest of the kernel from being vectorized, 
 because packing/unpacking values in and out of a vector register is often uneconomical.
By widening the available parallelism for the recurrence operators, our transformations let the entire kernel use vectorized instructions,
 so even non-recursive operators in a recurrence kernel benefit from our vectorization techniques.}

We also collect a set of 12 recurrences from dynamic programming problems on LeetCode.
\rev{These problems often exhibit diverse patterns, and no baselines are available.}
\autoref{tab:leetcode-throughput} shows the result of \vecrec{}--generated kernels against manual \CC{} implementations and unvectorized baselines.

\subsection{Case Study on Bioinformatic Recurrences}
\label{sec:case-study}

\begin{table}[t]
\centering
\caption{Comparing \vecrec{} with state-of-the-art vectorized libraries on sequence alignment algorithms of matrix size $2048 \times 2048$. The speedup is measured against the best-performing baselines. Ksw2 is run on Apple M2 via \texttt{sse2neon}. \textsuperscript{$\dagger$}For WFA2, runtime is proportional to sequence similarity and the 85\% similarity reports WFA2's speedup over \vecrec{}.}
\label{tab:biothroughput}
\resizebox{0.7\textwidth}{!}{%
\begin{tabular}{ll|rr|rr|rr}
\toprule
 &  & \multicolumn{2}{c}{Skylake} & \multicolumn{2}{c}{Zen 5} & \multicolumn{2}{c}{Apple M2} \\
\cmidrule(lr){3-4} \cmidrule(lr){5-6} \cmidrule(lr){7-8}
Benchmark & Implementation & Melem/s & \(\times\) & Melem/s & \(\times\) & Melem/s & \(\times\) \\
\midrule
Edit Distance & edlib & 4677.2 & -- & 9615.0 & -- & 7540.3 & -- \\
 & \textbf{VecRec} & 3690.6 & 0.79 & 12061.1 & 1.25 & 3030.8 & 0.40 \\
\midrule
Needleman--Wunsch & parasail\_scan & 2960.1 & -- & 5829.0 & -- & 1903.5 & -- \\
 & WFA2 (65\%) & 1417.9 & -- & 2795.4 & -- & 2113.1 & -- \\
 & parasail\_striped & 1065.9 & -- & 2328.2 & -- & 1273.3 & -- \\
 & parasail\_diag & 610.3 & -- & 2008.4 & -- & 1010.5 & -- \\
 & ksw2\_gg2\_sse & 1189.2 & -- & 2302.0 & -- & 1751.8 & -- \\
 & seqan3 & 172.5 & -- & 288.3 & -- & 226.8 & -- \\
 & \textbf{VecRec} & 3676.5 & 1.24 & 12002.0 & 2.06 & 3035.9 & 1.44 \\
 & WFA2 (85\%)\textsuperscript{$\dagger$} & 6124.0 & 1.67 & 11124.1 & 0.93 & 9859.8 & 3.25 \\
\midrule
Smith--Waterman & parasail\_striped & 2611.8 & -- & 5246.8 & -- & 2220.3 & -- \\
 & parasail\_scan & 2750.3 & -- & 4503.9 & -- & 1604.3 & -- \\
 & parasail\_diag & 587.5 & -- & 1869.7 & -- & 915.4 & -- \\
 & seqan3 & 158.0 & -- & 253.1 & -- & 200.9 & -- \\
 & \textbf{VecRec} & 2863.5 & 1.04 & 8150.4 & 1.55 & 2276.9 & 1.03 \\
\midrule
Affine Gap & parasail\_scan & 3023.9 & -- & 5789.0 & -- & 1934.4 & -- \\
 & parasail\_striped & 1007.4 & -- & 2187.2 & -- & 1284.1 & -- \\
 & WFA2 (65\%) & 767.5 & -- & 1581.6 & -- & 1191.3 & -- \\
 & parasail\_diag & 617.0 & -- & 1994.0 & -- & 1025.8 & -- \\
 & ksw2\_extd2\_sse & 1456.9 & -- & 2756.8 & -- & 1809.2 & -- \\
 & ksw2\_extz2\_sse & 1877.5 & -- & 3544.4 & -- & 2864.3 & -- \\
 & ksw2\_gg2\_sse & 1195.9 & -- & 2170.5 & -- & 1892.5 & -- \\
 & \textbf{VecRec} & 4120.0 & 1.36 & 10835.0 & 1.87 & 3025.7 & 1.06 \\
 & WFA2 (85\%)\textsuperscript{$\dagger$} & 4279.7 & 1.04 & 8241.7 & 0.76 & 7543.7 & 2.49 \\
\bottomrule
\end{tabular}
}
\end{table}


Sequence alignment is built on four core recurrences that share the same basic structure: Edit Distance, Needleman--Wunsch, Smith--Waterman, and Affine Gap.
Among them, Affine Gap is the most widely used and most complex~\cite{mafft2008, gotoh1982}, consisting of three mutually recursive recurrences.
These recurrences take time quadratic in the sequence length, and genetic data spans billions of characters~\cite{striped, wfa, edlib, parasail}, so they are often the primary bottleneck in genomics pipelines~\cite{minimap2,bwa-mem,bowtie2,star}.

\paragraph*{Runtime comparison}

We implement all four recurrences in \vecrec{} and compare them with existing libraries that use hand-tuned optimizations. 
Our baselines include ksw2, which compresses the recurrence using score differences~\citep{minimap2, Suzuki2018IntroducingDR}, parasail's \texttt{striped} and \texttt{scan} strategies, which vectorize cells from the previous row, parasail's \texttt{diagonal} strategy, which vectorizes independent cells along an antidiagonal~\citep{parasail}, and seqan3,  an optimized scalar implementation~\citep{seqan3}. 
We also compare against WFA~\citep{wfa}, which is algorithmically distinct and has performance proportional to sequence similarity, and Edlib, which is specialized to edit distance and uses bit-packing to compute multiple cells within a single machine word~\citep{edlib}.



\autoref{tab:biothroughput} shows the evaluation results.
Vectorized methods trade parallelism against memory locality. Ksw2 and \texttt{diagonal} vectorize fully independent antidiagonals at the cost of scattered memory access. \texttt{Striped} and \texttt{scan} keep access contiguous within a row but don't resolve the column's serial dependency. Using dilation, \vecrec{} can both parallelize the serial dependency and preserve memory locality, providing up to \(2.4\times\) speedup over these libraries.
On the other hand, edlib and WFA can outperform \vecrec{} under specific conditions. Edlib's bitpacking optimization applies only to edit distance. WFA is faster at high sequence similarity and with small scoring parameters, but it requires scores to be strictly positive and does not support Smith-Waterman. In practice, however, sequences are often divergent~\citep{minimap2,Suzuki2018IntroducingDR,nobrega2004}, and scores can be large and negative~\citep{blosum,pam}.

\paragraph*{Optimizing end-to-end pipelines}
\begin{wraptable}[13]{r}{0.45\textwidth}
\caption{Smith--Waterman runtime for $N\times N$ matrices, comparing untiled code, Reptile's best tiling, and \vecrec{}. Tile sizes were $4$--$512$ with the best being col 16, col 16, col 16, col 16.}
\label{tab:reptile-vs-vecrec-sw}
\centering
\small
\begin{tabular}{lrrrr}
\toprule
 & $1024$ & $2048$ & $4096$ & $8192$ \\
\midrule
Untiled (ms) & 1.419 & 4.996 & 19.96 & 79.65 \\
Reptile (ms) & 0.947 & 3.788 & 15.15 & 60.56 \\
\vecrec{} (ms) & 0.393 & 1.508 & 5.871 & 23.30 \\
\midrule
Speedup & $2.41\times$ & $2.51\times$ & $2.58\times$ & $2.60\times$ \\
\bottomrule
\end{tabular}
\end{wraptable}
Genomics pipelines use sequence alignment recurrences in performance-critical inner loops~\citep{bowtie2, bwa-mem, star, minimap2}. Minimap2, a widely used long-read alignment tool, issues thousands to millions of small affine-gap recurrence calls in parallel via its alignment library ksw2~\citep{minimap2}.

Ksw2 exposes several alignment modes trading accuracy for speed: exact alignment, which computes the full recurrence and guarantees the optimal score, and approximations such as early termination and reduced-precision scores, which users select based on their workload's error tolerance~\citep{minimap2}. The approximations are tied to ksw2's reformulated recurrence and antidiagonal traversal, so \vecrec{} cannot support them directly; instead, it serves as a drop-in accelerator for the exact mode.

We compare \vecrec{} with ksw2 on minimap2's exact mode.
Exact computation is useful for noisy reads\footnote{A DNA read is a fragment of DNA sequence output by a sequencer, and a noisy read is a read with a high error rate.} such as Oxford Nanopore Technologies (ONT) long reads~\citep{minimap2}. 
On ONT datasets, minimap2 integrated with \vecrec{} achieves up to $1.28\times$ speedup over stock Minimap2 (\autoref{tab:ont-vecrec-speedup}).



\begin{table}[t]
\caption{
End-to-end alignment time on ONT workloads with approximate maximum, banding, and z-drop flags disabled. \texttt{ecoli-ultralong} uses ultra-long \textit{E. coli} reads, \texttt{ecoli-strains} uses \textit{E. coli} strain sequences, and \texttt{pa-strains} uses \textit{Pseudomonas aeruginosa} strain sequences. These results were obtained on the Zen 5 microarchitecture.
}
\label{tab:ont-vecrec-speedup}
\centering
\resizebox{0.6\textwidth}{!}{
\small{
\begin{tabular}{lrrr}
\toprule
Dataset & Minimap2 (ksw2) & Minimap2 (\vecrec{}) & Speedup \\
\midrule
\texttt{ecoli-ultralong} & 58{,}235 ms & 45{,}624 ms & 1.28$\times$ \\
\texttt{ecoli-strains} & 5{,}226 ms & 4{,}183 ms & 1.25$\times$ \\
\texttt{pa-strains} & 1{,}898 ms & 1{,}761 ms & 1.08$\times$ \\
\bottomrule
\end{tabular}
}
}
\end{table}

\paragraph*{Comparison to Reptile~\citep{reptile}}
\looseness=-1 We compare \vecrec{} against Reptile, a tiling framework for recurrences. \vecrec{} outperforms Reptile by $2.37\times$, because Reptile generates only scalar code (\autoref{tab:reptile-vs-vecrec-sw}). Reptile also lets an external kernel replace its inner loop as a black box, so we composed it with \vecrec{} for sequence alignment, but this yielded no additional speedup. Reptile's optimal tile sizes are small and is covered by the unrolling \vecrec{} already applies.

\subsection{Understanding Vectorization Schedules}
\label{sec:schedules}

\begin{table}[t]
\caption{Vectorization schedules for the second-order filter
in float32,
with poles at $r e^{\pm i\theta}$ for a fixed $\theta$.
Speedup is over the scalar baseline.
Error is measured as the relative $L_2$ distance from a scalar float64 evaluation
on $2^{22}$ uniform random inputs, measured on Zen 5;
$\infty$ means the kernel overflows.}
\label{tab:schedules}
\centering
\resizebox{0.75\textwidth}{!}{
\small{%
\begin{tabular}{lrrr|rrr}
\toprule
Schedule & \multicolumn{3}{c|}{Speedup} & \multicolumn{3}{c}{Relative $L_2$ error} \\
 & Skylake & Zen 5 & Apple M2 & $r=0.9$ & $r=0.99$ & $r=0.999$ \\
\midrule
Scalar & 1.00$\times$ & 1.00$\times$ & 1.00$\times$ & $3.6\!\times\!10^{-7}$ & $1.2\!\times\!10^{-6}$ & $3.7\!\times\!10^{-6}$ \\
Dilation & 3.99$\times$ & 8.66$\times$ & 3.10$\times$ & $9.4\!\times\!10^{-8}$ & $1.7\!\times\!10^{-7}$ & $9.9\!\times\!10^{-7}$ \\
Dilation + Delay & 4.05$\times$ & 8.08$\times$ & 3.13$\times$ & $9.4\!\times\!10^{-8}$ & $\infty$ & $\infty$ \\
Fast associative scan & 1.76$\times$ & 3.54$\times$ & 2.93$\times$ & $4.4\!\times\!10^{-7}$ & $2.0\!\times\!10^{-5}$ & $1.1\!\times\!10^{-4}$ \\
Path-based decomp. & 1.74$\times$ & 3.22$\times$ & 2.82$\times$ & $1.0\!\times\!10^{-6}$ & $\infty$ & $\infty$ \\
\bottomrule
\end{tabular}

}
}
\end{table}

\begin{table}[t]
\caption{Fast associative scan vs.\ path-based decomposition on two applications based on time-varying recurrences. Speedup is measured over a scalar implementation.}
\label{tab:scan-vs-path}
\centering
\resizebox{0.6\textwidth}{!}{
\small
\begin{tabular}{l|rr|rr|rr}
\toprule
Benchmark & \multicolumn{2}{c}{Skylake} & \multicolumn{2}{c}{Zen 5} & \multicolumn{2}{c}{Apple M2} \\
 & Scan & Path & Scan & Path & Scan & Path \\
\midrule
Continued Fraction & \textbf{2.29$\times$} & 2.00$\times$ & \textbf{3.05$\times$} & 2.88$\times$ & 2.72$\times$ & \textbf{3.22$\times$} \\
Welford Variance & 1.81$\times$ & \textbf{2.33$\times$} & 2.41$\times$ & \textbf{3.78$\times$} & 1.68$\times$ & \textbf{2.76$\times$} \\
\bottomrule
\end{tabular}

}
\end{table}

\rev{
To understand the characteristics of individual transformations, we vectorize one representative second-order filter with each of them. \autoref{tab:schedules} shows their achieved speedup and accuracy. We measure the accuracy using three different configurations of the radius $r$ of the filter's poles. A larger radius $r$ makes the filter harder to evaluate accurately.

The result shows that, when applicable, dilation (possibly with delay) achieves the best speedup compared to other more general transformations.
Dilation makes the kernel \emph{more} accurate than scalar evaluation.
While dilation + delay may provide additional speedups, it can become numerically unstable as $r$ gets closer to $1$.

We also evaluate the two general transformations on two time-varying applications (\autoref{tab:scan-vs-path}).
Neither dominates the other, so the best schedule can only be determined through run-time profiling.


}

\section{Related Work}
\label{sec:related}

Several lines of work inspired this paper.
Parallel algorithms for associative scans date back to Kogge--Stone adders~\citep{kogge-stone,stone} and were popularized by \citet{BlellochTR90}. Our fast associative scan (\autoref{sec:assoc-scan}) is essentially a reformulation of recursive doubling \citep{kogge-stone,BlellochTR90} in the recurrence algebra.
\rev{To turn it into efficient vectorized instructions, however,
we propose a hardware-efficient scheme that interleaves the computation of vector lanes.}

In digital signal processing (DSP), \citet{simd-dsp-recursive-filters} applies \citet{BlellochTR90}-style parallelism in digital signal processors, and
 \citet{vectoring-iir-filters} vectorizes IIR filters by unrolling the recurrence definitions, i.e., delaying.
Our dilation for ring recurrences is known as scattered look-ahead in DSP literature~\citep{scattered-lookahead};
 we have shown the algebraic identities behind it.
Concurrent work by~\citet{ma2026findingfastfilters} also applies scattered look-ahead for IIR vectorization. 

In tropical semirings, \citet{jump-flooding} described jump flooding for parallelizing distance transforms.
This corresponds to our dilation transformation for idempotent recurrences (\autoref{lem:idem-dilation}).
\rev{
Our dilation transformation slightly generalizes jump flooding
 by allowing different taps to be dilated by different amounts.
For instance, 
 a variant of the distance transform used in practice is $F = \operatorname{Rec}\left(\left[\begin{smallmatrix}\infty,1,\infty\\ 1,\sqrt{2},\sqrt{5}\\ \infty, \sqrt{5},\infty\end{smallmatrix}\right], G\right)$. To compute a $2\times 2$ grid in parallel, we can let the two taps with coefficients $\sqrt{5}$ stay undilated and only dilate the rest, leading to a recurrence kernel $\left[\begin{smallmatrix}\infty,\infty,2\\ \infty,\infty,\sqrt{5}\\ 2, \sqrt{5},2\sqrt{2}\end{smallmatrix}\right]$. This is more efficient than classical jump flooding, which dilates every term.
}

\rev{
Beyond the generalizations to prior work,
 we treat vectorization schemes as generic transformations over algebraic structures (e.g., rings and tropical semirings),
 so they are not confined to a particular domain:
 jump flooding, for instance, had never been applied to bioinformatics or many other DP problems, yet we show that it achieves significant speedups there.
We also build a compiler to show that these approaches yield efficient SIMD vectorization,
 whereas prior work largely targeted DSP design or multicore parallelism.
}

In signal and image processing, prior work has developed parallel algorithms for computing time-invariant recursive filters \citep{gpu-summed-area}.
We draw inspiration from \textsf{RecFilter}~\citep{recfilter}, a compiler that automatically generates high-performance parallel implementations of multi-dimensional time-invariant IIR filters.
Their approaches to multi-core parallelism, however, do not directly extend to vectorization, because they require simultaneously loading from non-contiguous dimensions.
\rev{Multi-core parallelism for recurrences (outer-loop) is orthogonal to this work (inner-loop). In fact, they complement each other: \autoref{tab:ont-vecrec-speedup} shows that minimap2, a multi-threaded alignment library, is further sped up by \vecrec{}'s vectorization.}

Recent domain-specific compilers for recurrences include
 \textsc{RecUMA}~\citep{recuma}, which compiles recurrences to efficient implementations over dense and sparse arrays,
 and \textsc{Reptile}~\citep{reptile}, which explores tiling strategies.
They handle a more general class of recurrences, including ones that aggregate over \emph{all} previous values, for which auto-vectorization suffices.
We focus on fixed-order recurrences, which require further techniques to vectorize efficiently.

Our recurrence algebra is inspired by recent semiring-based database languages~\citep{datalogo,provenance-semirings,dbsp}.
Interestingly, DBSP~\citep{dbsp}, a language for incremental computation, is also inspired by signal processing and uses terminology from  Z-transform.
Unlike these works, our algebra is a semimodule rather than a semiring, because convolution does not apply between signals.
Similar to our work, \citet{generalized-convolution} casts convolution in terms of semirings and semimodules, unifying signal processing, polynomial multiplication, probability, and language recognition.

Hand-optimized sequence alignment libraries use two primary vectorization paradigms. Row-based approaches (e.g., Parasail's \texttt{striped} and \texttt{scan}~\citep{parasail,striped}) vectorize the dependency on the previous row, whereas antidiagonal approaches (e.g., Parasail's \texttt{diagonal}, ksw2~\citep{Suzuki2018IntroducingDR}, and WFA~\citep{wfa}) vectorize along independent diagonals. To maximize performance, these libraries employ specialized optimizations: ksw2 leverages score differences for \texttt{int8} compression, WFA uses search-based recurrences to accelerate highly similar sequences, and Edlib~\citep{edlib} packs into machine words. While these frameworks are restricted to specific scoring strategies, 
\vecrec{} is scoring-agnostic and supports diverse alignment and score variants. End-to-end genomic pipelines such as minimap2~\citep{minimap2}, BWA-MEM~\citep{bwa-mem}, Bowtie2~\citep{bowtie2}, and STAR~\citep{star} embed these libraries to process large-scale sequence data.
Filtr~\citep{filtr} is a recent compiler for bioinformatics recurrences that matches or beats the performance of hand-tuned implementations.

Our work is also inspired by \citet{hardboiled},
 who sped up recursive filters by first applying scattered lookahead and then accelerating the resulting convolution using Tensor Cores.
In a similar vein,
 our techniques may be adapted to other hardware, including SIMT GPUs and matrix accelerators.

\section{Conclusion and Future Work}

In this paper,
 we present the recurrence algebra over semirings,
 which captures a wide variety of recurrences in the wild,
 we show a set of identities and transformations that effectively vectorize them,
 and we build the \vecrec{} compiler based on these ideas.
Experiments show that the compiler-generated kernels can be an order of magnitude faster than non-vectorized implementations and are competitive with expert-optimized vectorized libraries.

This paper focuses on linear recurrences over semirings,
 so natural next steps are nonlinear recurrences,
 such as the Mandelbrot set or Floyd-Steinberg dithering,
 and recurrences not over semirings,
 as in certain LeetCode problems.
On the implementation side,
 we would like to apply our framework to the GPU's SIMT model.
\rev{\vecrec{} currently supports only transformations expressible in the recurrence algebra,
which are shift-invariant (e.g., $f(x)$ and $f(x-1)$ are computed using the same scheme),
yet many transformations in domain-specific libraries are not,
such as bitpacking in edit distance and parasail's wavefront parallelism.
Future work is thus a compiler that supports and combines different vectorization and multi-core parallelism schemes, as well as other transformations like loop reordering~\citep{recuma} and multithreading.
}

\begin{acks}
We thank the anonymous reviewers for their valuable feedback, Shoaib Kamil and Dan Suciu for their feedback on early versions of this work, and Shiv Sundram and Muhammad Usman Tariq for helping with the REPTILE experiment.
This material is based upon work supported by the National Science Foundation under Grant Nos. 2232339 and 2312195. Any opinions, findings, and conclusions or recommendations expressed in this material are those of the author(s) and do not necessarily reflect the views of the National Science Foundation.
\end{acks}

\section{Disclosure of Use of AI in Research}

\looseness=-1 AI tools assisted with implementing the compiler, particularly features needed for the bioinformatics-related case studies, developing the evaluation harness, and implementing C++ scalar baselines for the LeetCode and signal processing problems. AI tools were used for formatting and proofreading prose and for producing some initial paragraph drafts; these drafts were subsequently heavily edited by the authors.

\section*{Data Availability Statement}
The source code of the \vecrec{} compiler is available at \url{https://github.com/yihozhang/vec-rec}. The artifact additionally includes the evaluation harness necessary to reproduce the experiment in this paper, and it is available at \url{https://doi.org/10.5281/zenodo.23330549} \citep{vecrec-artifact}.

\newpage

\bibliographystyle{ACM-Reference-Format}
\bibliography{main}

\newpage 
\appendix

\section{Proof of theorems concerning the recurrence algebra}
\label{app:kernel-proofs}

\begin{lemma}
    In the arithmetic ring over real numbers,
    any one-dimensional time-invariant kernel $A$ of order $n$ can
    be factored as $A = A_1\otimes \ldots\otimes A_k$ where each $A_i$ 
    has order less than or equal to 2.
\end{lemma}
\begin{proof}
We can write $A$ as coefficients of a generating function
$p_A(x) = \sum_{i=0}^n a_i x^{i}$, where $a_i$ is the coefficient of the $i$-th tap of $A$.
The fundamental theorem of algebra states that $p_A$ can be factored into 
a product of real polynomials of degree at most 2,
 so we have $p_A(x) = p_{A_1}(x)\cdot \ldots \cdot p_{A_k}(x)$,
 where each $p_{A_i}$ is either linear ($ax+b$) or quadratic ($ax^2+bx+c$).
\end{proof}

\section{Proof of theorems concerning recurrences}

\begin{lemma}[Equivalent definitions of \rec{}]
    Suppose $A$ is causal.
    $\bigoplus_{i\geq 0} A^i\otimes G$ is well-defined, and 
    $F = \bigoplus_{i\geq 0} A^i\otimes G$ iff $F=A\otimes F\oplus G$.
\end{lemma}

\begin{proof}
We assert $\semof{A^k}_{\veci}=\algzero$ if the L1 norm $|\veci|_1$ is smaller than $k$ for any $k$.
This can be proved by induction on $k$
Therefore, for a given index $\vecn$,
 we can pick any $k$ satisfying $|\vecn|_1 \geq k$,
 and $\semof{A^{k}\otimes F}(\vecn)=\algzero$.

Fix any such $k$.
Then $\semof{\bigoplus_{i\geq 0} A^i\otimes G}(\vecn)$
 is well-defined and equal to $\semof{\bigoplus_{i=0}^{k} A^i\otimes G}(\vecn)$.
Therefore, the semantics of the unbounded sum
 $\bigoplus_{i\geq 0} A^i\otimes G$ is well-defined for any index $\vecn$.

($\Leftarrow$) Suppose $F=A\otimes F\oplus G$. 
For any $k$, we have
\begin{align*}
F&=A\otimes F\oplus G\\
&=A\otimes (A\otimes F\oplus G)\oplus G\\
&= A^2\otimes F\oplus A\otimes G\oplus G\\
&\;\;\vdots\\
&=A^{k}\otimes F\;\oplus\;\bigoplus_{i=0}^{k-1} A^i\otimes G
\end{align*}

For any $\vecn$, we can choose a $k$ with $k\cdot\vecone>\vecn$
 so that 
 \[\semof{F}(\vecn)=\semof{A^{k}\otimes F}(\vecn)\algplus \semof{\bigoplus_{i=0}^{k-1} A^i\otimes G}(\vecn)=
 \semof{\bigoplus_{i=0}^{k-1} A^i\otimes G}(\vecn)=\semof{\bigoplus_{i\geq 0} A^i\otimes G}(\vecn).
 \]

($\Rightarrow$) Conversely, if $F=\bigoplus_{i\geq 0} A^i\otimes G$ then
\begin{align*}
A\otimes F\oplus G
&=A\otimes \left(\bigoplus_{i\geq 0} A^i\otimes G\right)\oplus G\\
&=\bigoplus_{i\geq 0} A^{i+1}\otimes G\;\oplus\;A^0\otimes G\\
&= \bigoplus_{i\geq 0} A^i\otimes G = F.
\end{align*}
\end{proof}

\subsection{Algebraic Identities over Any Semiring}

\begin{lemma}[Unrolling]
For kernel $A$ and signal $G$ over any semiring,
\[
    \rec(A, G) = \rec\left(A^k,\ \bigoplus_{i=0}^{k-1} A^i\otimes G\right).
\]
\end{lemma}
\begin{proof}
\begin{align*}
    \rec(A, G) &= \bigoplus_{j\geq 0} A^j\otimes G\\
               &= \bigoplus_{j\geq 0} \bigoplus_{i=0}^{k-1}A^{j\cdot k+i}\otimes G\\
               &= \bigoplus_{j\geq 0} \left(A^{j\cdot k}\otimes \bigoplus_{i=0}^{k-1}A^i\otimes G \right)\\
               &= \rec\left(A^k,\ \bigoplus_{i=0}^{k-1} A^i\otimes G\right). \qedhere
\end{align*}
\end{proof}

\begin{lemma}[Commuting recurrence with convolution]
For time-invariant kernels $A$ and $B$ and signal $G$ over any semiring,
\[
    \rec(A, B\otimes G) = B\otimes \rec(A, G).
\]
\end{lemma}
\begin{proof}
\begin{align*}
\rec(A, B\otimes G)
 &= \bigoplus_{k=0}^\infty A^k\otimes B\otimes G\\
 &= \bigoplus_{k=0}^\infty B\otimes A^k\otimes G\\
 &= B\otimes\rec(A, G). \qedhere
\end{align*}
\end{proof}

\begin{lemma}[Commuting recurrences]
For time-invariant kernels $A$ and $B$ and signal $G$ over any semiring,
\[
    \rec(A, \rec(B, G)) = \rec(B, \rec(A, G)).
\]
\end{lemma}
\begin{proof}
\begin{align*}
    \rec(A, \rec(B, G))
    &= \bigoplus_{i\geq 0}A^i\otimes \left(\bigoplus_{j\geq 0}B^j\otimes G\right)\\
    &= \bigoplus_{i\geq 0} \bigoplus_{j\geq 0}\left(A^i\otimes B^j\otimes G\right)\\
    &= \bigoplus_{j\geq 0} \bigoplus_{i\geq 0}\left(A^i\otimes B^j\otimes G\right)\\
    &= \rec(B, \rec(A, G)). \qedhere
\end{align*}
\end{proof}

\subsection{Algebraic Identities over Rings}

\begin{lemma}[Inverse of Recurrence]
For kernel $A$ and signal $G$ over a ring,
\[
    (I\ominus A)\otimes \rec(A,G) = G.
\]
\end{lemma}
\begin{proof}
Let $F = \rec(A, G)$, so $F = A\otimes F\oplus G$. Then
\begin{align*}
(I\ominus A)\otimes F &= F\ominus A\otimes F = G. \qedhere
\end{align*}
\end{proof}

\begin{lemma}[Fusion of Stacked Ring Recurrences]
For kernels $A$ and $B$ and signal $G$ over a ring,
\[
    \rec(A, \rec(B, G)) = \rec(A\oplus B\ominus B\otimes A,\,G).
\]
\end{lemma}
\begin{proof}
Let $F = \rec(A, \rec(B, G))$, so $F = A\otimes F\oplus\rec(B, G)$ and hence
$(I\ominus A)\otimes F = \rec(B, G)$. Therefore
\begin{align*}
(I\ominus A)\otimes F &= B\otimes (I\ominus A)\otimes F\oplus G\\
(I\ominus B)\otimes(I\ominus A)\otimes F &= G\\
(I\ominus A\ominus B\oplus B\otimes A)\otimes F &= G\\
F &= \rec(A\oplus B\ominus B\otimes A,\,G). \qedhere
\end{align*}
\end{proof}

\begin{lemma}[Fusion of Parallel Ring Recurrences]
For time-invariant kernels $A$ and $B$ and signals $G$ and $H$ over a ring,
\[
    \rec(A, G)\oplus\rec(B, H) = \rec\bigl(A\oplus B\ominus A\otimes B,\,(I\ominus B)\otimes G\oplus(I\ominus A)\otimes H\bigr).
\]
\end{lemma}
\begin{proof}
Let $F_1 = \rec(A, G)$ and $F_2 = \rec(B, H)$, so $(I\ominus A)\otimes F_1 = G$ and $(I\ominus B)\otimes F_2 = H$.
Setting $F = F_1\oplus F_2$:
\begin{align*}
(I\ominus A)\otimes(I\ominus B)\otimes F &= (I\ominus B)\otimes G\oplus(I\ominus A)\otimes H\\
(I\ominus A\ominus B\oplus A\otimes B)\otimes F &= (I\ominus B)\otimes G\oplus(I\ominus A)\otimes H\\
F &= \rec\big(A\oplus B\ominus A\otimes B,\,(I\ominus B)\otimes G\oplus(I\ominus A)\otimes H\big). \qedhere
\end{align*}
\end{proof}

\begin{lemma}[Dilation with Additive Inverse]
For time-invariant kernels $A$ and $B$ and signal $G$ over a ring,
\[
    \rec(A\oplus B,G) = \rec\bigl(A\oplus A\ominus A^2\oplus B^2,\,(I\ominus A\oplus B)\otimes G\bigr).
\]
\end{lemma}
\begin{proof}
Let $F = \rec(A\oplus B, G)$, so $(I\ominus A\ominus B)\otimes F = G$. Multiplying by $(I\ominus A\oplus B)$:
\begin{align*}
(I\ominus A\oplus B)\otimes(I\ominus A\ominus B)\otimes F &= (I\ominus A\oplus B)\otimes G\\
\big((I\ominus A)^2\ominus B^2\big)\otimes F &= (I\ominus A\oplus B)\otimes G\\
(I\ominus A\ominus A\oplus A^2\ominus B^2)\otimes F &= (I\ominus A\oplus B)\otimes G\\
F &= \rec\big(A\oplus A\ominus A^2\oplus B^2,\,(I\ominus A\oplus B)\otimes G\big). \qedhere
\end{align*}
\end{proof}

\begin{lemma}[Decomposition of Squared Kernels]
For kernel $A$ and signal $G$ over a ring,
\[
    \rec(A^2, G) = \rec(A, \rec(\ominus A, G)).
\]
\end{lemma}
\begin{proof}
By Lemma~\ref{lem:flatten} with $B := \ominus A$:
\begin{align*}
\rec(A, \rec(\ominus A, G))
 &= \rec\big(A\oplus(\ominus A)\ominus(\ominus A)\otimes A,\,G\big)\\
 &= \rec(\algzero\oplus A^2,\,G)\\
 &= \rec(A^2,\,G). \qedhere
\end{align*}
\end{proof}

\begin{lemma}[Change of Basis]
For kernels $A$ and $B$ and signal $G$ over a ring,
\[
    \rec(A, G)= \rec(A\ominus B\otimes(I\ominus A),\ (I\oplus B)\otimes G).
\]
\end{lemma}
\begin{proof}
Let $F = \rec(A, G)$, so $(I\ominus A)\otimes F = G$. Multiplying by $(I\oplus B)$:
\begin{align*}
(I\oplus B)\otimes(I\ominus A)\otimes F &= (I\oplus B)\otimes G\\
(I\ominus A\oplus B\ominus B\otimes A)\otimes F &= (I\oplus B)\otimes G\\
F &= \rec\big(A\ominus B\otimes(I\ominus A),\,(I\oplus B)\otimes G\big). \qedhere
\end{align*}
\end{proof}

\subsection{Algebraic Identities over Idempotent Semirings}

\begin{lemma}[Dilation with Idempotence]
For time-invariant kernels $A$ and $B$ and signal $G$ over an idempotent semiring,
\[
    \rec(A\oplus B, G) = \rec(A\oplus B^2,\ (I\oplus B)\otimes G).
\]
\end{lemma}
\begin{proof}
Since $A$ and $B$ are time-invariant and $\oplus$ is idempotent, the binomial
expansion collapses to
\[(A\oplus B)^k = \bigoplus_{i+j=k} A^i\otimes B^j,\]
and therefore
\[\rec(A\oplus B, G) = \bigoplus_{k>0}(A\oplus B)^k\otimes G = \bigoplus_{i,j\geq 0} A^i\otimes B^j\otimes G.\]
Reindexing $j$ by parity:
\begin{align*}
    \rec(A\oplus B, G)
      &= \bigoplus_{i,j\geq 0} A^i\otimes B^{2j}\otimes G\ \oplus\ \bigoplus_{i,j\geq 0} A^i\otimes B^{2j+1}\otimes G\\
      &= \bigoplus_{i,j\geq 0} A^i\otimes B^{2j}\otimes (I\oplus B)\otimes G\\
      &= \bigoplus_{k>0} (A\oplus B^2)^k\otimes (I\oplus B)\otimes G\\
      &= \rec\left(A\oplus B^2,\ (I\oplus B)\otimes G\right). \qedhere
\end{align*}
\end{proof}

\begin{lemma}[Fusion of Idempotent Recurrences]
For time-invariant kernels $A$ and $B$ and signal $G$ over an idempotent semiring,
\[
    \rec(A, \rec(B, G)) = \rec(A\oplus B, G)
\]
\end{lemma}
\begin{proof}
Using the same expansion as in the proof of Lemma~\ref{lem:idem-dilation}:
\begin{align*}
    \rec(A\oplus B, G)
      &= \bigoplus_{i,j\geq 0} A^i\otimes B^j\otimes G\\
      &= \bigoplus_{i>0} A^i\otimes \left(\bigoplus_{j>0} B^j\otimes G\right)\\
      &= \bigoplus_{i>0} A^i\otimes \rec(B, G)\\
      &= \rec(A, \rec(B, G)). \qedhere
\end{align*}
\end{proof}

\section{Proof of path-based decomposition}



\begin{lemma}
Let $F$ be a recurrence defined by $F=\rec(A, G)$. We have
    \[
    F=
    G_p
    \oplus 
    \left(
    \bigoplus_{1\leq q\leq k}
    \bigoplus_{\substack{0\leq r< p\\ p\leq r+q }}
     B_r\otimes K_q\otimes 
    F
    \right)\]
\end{lemma}
\begin{proof}
\begin{align*}
F&=
    \bigoplus_{n\geq 0} B_n \otimes G\\
&=
    \left(\bigoplus_{0\leq n < p} B_n \otimes G\right)
    \oplus 
    \left(\bigoplus_{n \geq p} B_n \otimes G\right)
    \\
&=
    G_p
    \oplus 
    \left(\bigoplus_{n \geq p} 
    \bigoplus_{1\leq q\leq k} \bigoplus_{\substack{0\leq r< p\\ p\leq r+q\leq n }}B_r\otimes K_q\otimes B_{n-r-q}
    \otimes G
    \right)\\
&=
    G_p
    \oplus 
    \left(
    \bigoplus_{1\leq q\leq k}
    \bigoplus_{n \geq p} 
    \bigoplus_{\substack{0\leq r< p\\ p\leq r+q\leq n }}
     B_r\otimes K_q\otimes B_{n-r-q}
    \otimes G
    \right)
    \\
&=
    G_p
    \oplus 
    \left(
    \bigoplus_{1\leq q\leq k}
    \bigoplus_{\substack{0\leq r< p\\ p\leq r+q }}
    \bigoplus_{n \geq r+q} 
     B_r\otimes K_q\otimes B_{n-r-q}
    \otimes G
    \right)
    \\
&=
    G_p
    \oplus 
    \left(
    \bigoplus_{1\leq q\leq k}
    \bigoplus_{\substack{0\leq r< p\\ p\leq r+q }}
     B_r\otimes K_q\otimes 
    \bigoplus_{n' \geq 0} B_{n'} \otimes G
    \right)
    \\
&=
    G_p
    \oplus 
    \left(
    \bigoplus_{1\leq q\leq k}
    \bigoplus_{\substack{0\leq r< p\\ p\leq r+q }}
     B_r\otimes K_q\otimes 
    F
    \right)
\end{align*}
\end{proof}

\begin{lemma}
    For any $n, p$ where $p< n$,
\[
B_n=\bigoplus_{1\leq q\leq k} \bigoplus_{\substack{0\leq r< p\\ p\leq r+q\leq n }}{B_r}\otimes K_q\otimes {B_{n-r-q}}
\]
\end{lemma}
\begin{proof}
\begin{align*}
    B_n&=\bigoplus_{\substack{1\leq m\leq n\\a_1+\ldots+a_m=n\\\forall i.\ 1\leq a_i\leq k}} K_{a_1}\otimes \ldots\otimes K_{a_m}\\
&={\color{Burgundy}\bigoplus_{1\leq q\leq k}}
    \bigoplus_{\substack{
        1\leq j \leq m\leq n\\
        {\color{Burgundy} a_j=q}\\
        a_1+\ldots+a_m=n   \\
        a_1+\ldots+a_{j-1}\leq p\\
        {\color{Burgundy}a_1+\ldots+a_{j-1}+q> p}\\
        \forall i.\ 1\leq a_i\leq k
    }
} K_{a_1}\otimes \ldots \otimes K_q\otimes\ldots  \otimes K_{a_m}
\\
&=\bigoplus_{1\leq q\leq k} 
{\color{Burgundy}\bigoplus_{\substack{0\leq r< p\\ p\leq r+q\leq n }}}
    \bigoplus_{\substack{
        1\leq j \leq m\leq n\\
        a_j=q\\
        a_1+\ldots+a_m=n   \\
        {\color{Burgundy} a_1+\ldots+a_{j-1}= r}\\
        \forall i.\ 1\leq a_i\leq k
    }
} K_{a_1}\otimes \ldots \otimes K_q\otimes\ldots  \otimes K_{a_m} \\
&=\bigoplus_{1\leq q\leq k} \bigoplus_{\substack{0\leq r< p\\ p\leq r+q\leq n }}\left( 
{\color{Burgundy}
\bigoplus_{\substack{
        1\leq m \leq r\\
        a_1+\ldots+a_m=r   \\
        \forall i.\ 1\leq a_i\leq k
    }
} K_{a_1}\otimes \ldots \otimes K_{a_m}
}
\right)
\otimes K_q \otimes
\left(
{\color{Burgundy}
\bigoplus_{\substack{
        1\leq m \leq n-r-q\\
        a_1+\ldots+a_m=n-r-q   \\
        \forall i.\ 1\leq a_i\leq k
    }
} K_{a_1}\otimes \ldots \otimes K_{a_m}
}
\right) \\
&=\bigoplus_{1\leq q\leq k} \bigoplus_{\substack{0\leq r< p\\ p\leq r+q\leq n }}{\color{Burgundy}B_r}\otimes K_q\otimes {\color{Burgundy}B_{n-r-q}}
\end{align*}
\end{proof}

\begin{lemma}
    For any $p, l$ where $l<p$
\[G_p=
   G_l
   \oplus
   \left(
   \bigoplus_{1\leq q\leq k} 
   \bigoplus_{\substack{0\leq r< l\\ l\leq  r+q}}
   {B_r}\otimes K_q\otimes 
   G_{p-r-q}
   \right)\]
\end{lemma}
\begin{proof}
\begin{align*}
   G_p&=\bigoplus_{0\leq n < p} B_n \otimes G\\
   &=  
   \left(\bigoplus_{0\leq n < l} B_n \otimes G\right)
   \oplus 
   \left(
   \bigoplus_{l\leq n < p}
   \bigoplus_{1\leq q\leq k} 
   \bigoplus_{\substack{0\leq r< l\\ l\leq  r+q\leq n }}{B_r}\otimes K_q\otimes {B_{n-r-q}}\otimes G
   \right)
   \\
   &=  
   G_l
   \oplus
   \left(
   \bigoplus_{1\leq q\leq k} 
   \bigoplus_{\substack{0\leq r< l\\ l\leq  r+q}}
   {B_r}\otimes K_q\otimes 
   \bigoplus_{r+q\leq n < p}{B_{n-r-q}}\otimes G
   \right)
   \\
   &=  
   G_l
   \oplus
   \left(
   \bigoplus_{1\leq q\leq k} 
   \bigoplus_{\substack{0\leq r< l\\ l\leq  r+q}}
   {B_r}\otimes K_q\otimes 
   \bigoplus_{0\leq n' < p-r-q}{B_{n'}}\otimes G
   \right)
   \\
   &=
   G_l
   \oplus
   \left(
   \bigoplus_{1\leq q\leq k} 
   \bigoplus_{\substack{0\leq r< l\\ l\leq  r+q}}
   {B_r}\otimes K_q\otimes 
   G_{p-r-q}
   \right)
\end{align*}
\end{proof}

\section{\vecrec{}'s domain-specific language}

 \begin{figure}[t]
  \centering
  \begin{lstlisting}[mathescape=true,basicstyle=\ttfamily\footnotesize,
    classoffset=1,
    keywordstyle=\color{purple},
    alsoletter={<:},
    morekeywords={type, for,or,<:},
    classoffset=2,
    morekeywords={TimeInv,TimeVar,TimeInv2D,TimeVar2D,Num,Impulse,Var,Var2D,SignalVec,VectorAt,KernelMat,Repeater,Ith,KAdd,KConvolve,KNeg,SAdd,PointwiseMul,SNeg,Convolve,Convolve2D,Recurse,Recurse2D,WindowMap,ConvertLanes}]
type Kernel, Signal, Kernels, Signals <: Any
type Kernel1D, Kernel2D <: Kernel
type Signal1D, Signal2D <: Signal
type Mat<K> <: Kernels for K <: Kernel
type Vec<S> <: Signals for S <: Signal

TimeInv : [Number] -> Kernel1D;    TimeInv2D : [[Number]] -> Kernel2D
TimeVar : [Signal1D] -> Kernel1D;  TimeVar2D : [[Signal1D]] -> Kernel2D

Num : Number -> Signal1D;  Impulse : Number -> Signal1D
Var : String -> Signal1D;  Var2D   : String -> Signal2D

SignalVec : [S] -> Vec<S> for S <: Signal
VectorAt  : (Vec<S>, Int) -> S for S <: Signal
KernelMat : [[K]] -> Mat<K> for K <: Kernel

Repeater : (Signal2D -> Signal1D) -> Signal2D
         | (Vec<Signal2D> -> Vec<Signal1D>) -> Vec<Signal2D>
Ith      : (Signal2D, Int) -> Signal1D

KAdd, KConvolve : (K, K) -> K for K <: Kernel or K <: Kernels
KNeg            : K -> K      for K <: Kernel or K <: Kernels

SAdd, PointwiseMul 
     : (Signal1D, Signal1D) -> Signal1D
     | (Vec<Signal1D>, Vec<Signal1D>) -> Vec<Signal1D>
SNeg : Signal1D -> Signal1D
     | Vec<Signal1D> -> Vec<Signal1D>

Convolve, Recurse
            : (Kernel1D, Signal1D) -> Signal1D
            | (Mat<Kernel1D>, Vec<Signal1D>) -> Vec<Signal1D>
Convolve2D : (Kernel2D, Signal2D) -> Signal1D
            | (Mat<Kernel2D>, Vec<Signal2D>) -> Vec<Signal1D>
Recurse2D   : (Kernel2D, Signal1D) -> Signal2D
            | (Mat<Kernel2D>, Vec<Signal1D>) -> Vec<Signal2D>

WindowMap : (n: Number, [S], [[S; n]] -> S) -> S
          for S : Signal1D | Vec<Signal1D>

ConvertLanes : (T, Int) -> T for T <: Any
\end{lstlisting}
\caption{The domain-specific language used in \vecrec{}.}
\label{fig:reclang-syntax-appendix}
\end{figure}

\end{document}